\documentclass[11pt,twoside]{article}
  \newcommand{\draft}{true}
  \ifx\draft\undefined
  \else
  \fi
  \usepackage{epic,eepic,psfig,latexsym,amssymb}
  \newcommand{\beq}{\begin{equation}}
  \newcommand{\eeq}{\end{equation}}
  \title{Transition to Centrifugal Particle Motion in Rotating Drums}
  \author{Gerald H. Ristow\\[1ex]
          {\em Fachbereich Physik, Philipps--Universit\"at} \\
          {\em Renthof 6, D--35032 Marburg, Germany} }
  \date{\small (received June 3, 1998; revised June 22, 1998)}
\begin{document}

\maketitle

\begin{abstract}
The dynamics and the transition to the centrifugal regime are studied
analytically and numerically for particles in rotating drum. The importance of
the particle-wall friction coefficient is demonstrated by studying first the
motion of one non-rotating particle where three different regimes are found in 
the transition to the centrifugal motion. When a few rotating particles are
considered, they behave similarly to one non-rotating particle in the low
friction limit. A critical particle number is necessary to reach the
centrifugal regime for which an analytic expression is derived in the limit of
negligible inter-particle friction. 
\end{abstract}

\ifx\draft\undefined
  \vfill
\fi

\begin{flushleft}
\end{flushleft}

\ifx\draft\undefined
  \newpage
\fi

%%%%%%%%%%%%%%%%%%%%%%%%%%%%%%%%%%%%%%%%%%%%%%%%%%%%%%%%%%%%%%%%%%%%%%%%%%%%%%%
\section{Introduction}
Granular materials can show fascinating behaviour in many experimental
situations~\cite{jaeger96}. When placed in a slowly rotating drum, most 
particles undergo a solid body rotation by following the external motion and a
thin layer of intermittently flowing particles (avalanches) is forming close to
the surface. With increasing rotation speed, the separation time of avalanches
will decrease until no individual avalanches are detectable and a nearly
constant particle flow is found along the free surface~\cite{rajchenbach90}.
For even higher rotation speeds, the free surface does not stay flat but rather
deforms with increasing rotation speed. These deformations mostly start from
the lower boundary inwards~\cite{dury97d}. When the rotation speed is further
increased, the particle motion is dominated by the centrifugal force leading to
a ring formation close to the outer wall boundary~\cite{nityanand86}. In order
to understand the different flowing regimes, it is instructive to study the
particle dynamics on a microscopic level, i.e.\ on a particle basis. This is
especially helpful for studying different frictional forces~\cite{caponeri95}
which play an important role in many other technological problems as
well~\cite{bhushan95}. In a simplified view, one can look at the particles
undergoing a solid body rotation as sticking to their neighbours whereas the
particles in the fluidized surface layer undergo mostly sliding collisions. A
similar particle behaviour can be observed for a particle placed on a moving
plate and attached to one side wall via a spring. For low surface velocities, a
stick-slip motion is observed and above a critical velocity, the stick-slip
motion disappears and only steady sliding is observed~\cite{heslot94,elmer97}.
But this setup might not allow for long enough observation times in order to
investigate transient behaviour. We propose another setup, namely to place the
particle(s) in a rotating drum.

The paper is organized in the following way: in section~\ref{sec: one_part}, we
will sketch the system in mind, explain briefly an experimental realization and
derive the equation of motion for one non-rotating particle in the full-sliding
limit. In section~\ref{sec: numeric}, we will present a numerical model which
is not restricted to the full-sliding limit but is capable to explore the full
particle dynamics including the transition from sticking to sliding motion.
This will be compared in section~\ref{sec: results} to the analytic solution
and the transition to the {\em centrifugal regime} will be discussed. In 
section~\ref{sec: few_part}, the collective behaviour of a few rotating spheres
in the drum will be investigated and a connection to the one particle case can
be made. The conclusions will round off this paper.

\section{Setup and Theory for One Non-Rotating Particle}
\label{sec: one_part}
Different friction regimes for particle-wall contacts can be studied by looking
at the dynamics of a non-rotating particle placed in a rotating drum. The setup
is sketched in Fig.~\ref{fig: one_part1}a where the particle is approximated by
a sphere shown in grey. The drum rotates with a constant angular velocity
$\Omega$, gravity is pointing downwards and the angle $\varphi$ of the
particle's center of mass is measured from the vertical. The distance between
the drum's and the particle's center is $R$. In the corresponding experimental
system, the non-rotating particle was realized by glueing two spheres together
and the drum was slightly tilted in order to force the particle to slide along
one of the end caps~\cite{betat97}.

The equation of motion is one-dimensional as long as the particle stays in
contact with the wall of the drum, i.e. $R\equiv \textrm{constant}$. For full
sliding, $|\dot{\varphi}| < \Omega$, the equation  of motion reads
\begin{eqnarray}
  \frac{F}{m} =  R\ddot{\varphi} &=& -\underbrace{g \sin \varphi}_{F_s} + \,
                  \mu_w \underbrace{( g \cos \varphi +
                  R {\dot{\varphi}}^2 )}_{F_n} \nonumber \\[-1ex]
              & &   \label{eq: one_part1} \\[-1ex]
              &=& - g\, \sqrt{1 + \mu_w^2}\, \sin (\varphi - \varphi_0) + \mu_w 
	          R {\dot{\varphi}}^2 \nonumber
\end{eqnarray}
where $\varphi_0 \equiv \textrm{arctan}(\mu_w)$ and the Coulomb friction law was
used. Here $F_n$ denotes the normal force and $F_s$ the shear force. Having the
many particle program at hand, we used MD simulations and compared the results
for one non-rotating particle with the experiment and the theory. The angle 
$\varphi$ as function of time for a typical run is shown in Fig.~\ref{fig:
one_part1}b using a drum rotation speed of $\Omega=1$~Hz and the initial
conditions $\varphi(0)=\dot{\varphi}(0)=0$.
\ifx\draft\undefined
\else
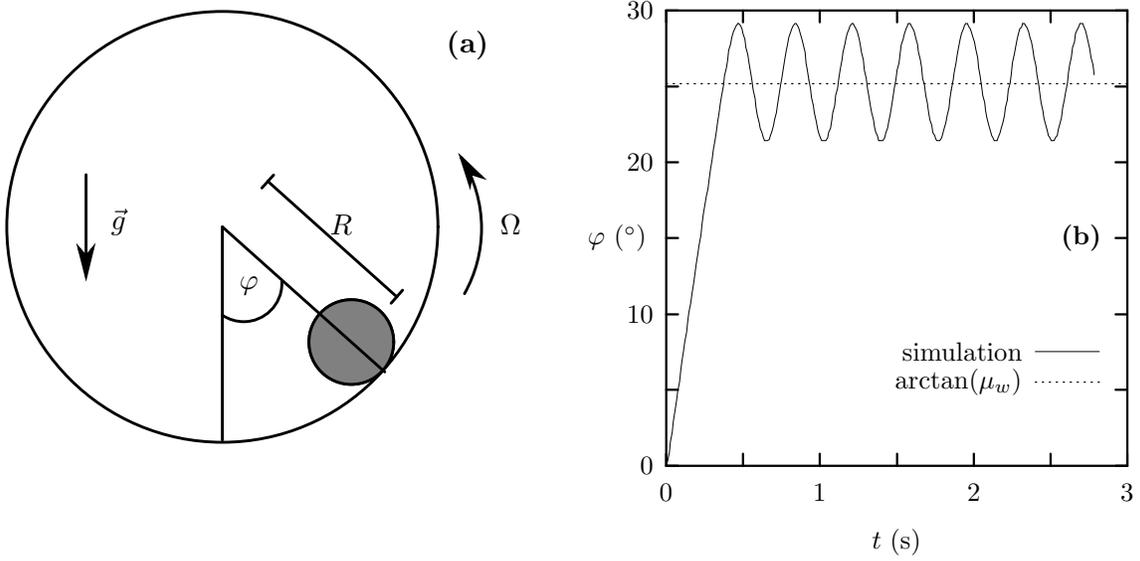
\begin{figure}[t]
  \hbox{\hspace{-0.7cm}\setlength{\unitlength}{0.0006in}
%{\renewcommand{\dashlinestretch}{30}
\begin{picture}(5500,3200)(-300,-850)

\Thicklines

\put(1927,1927){\circle{3794}}
\put(3062,912){\circle{750}}
\put(3062,912){\shade\circle{750}}

\blacken\path(4207.565,2367.785)(4029.000,2539.000)(4105.986,2303.897)(4118.443,2396.789)(4207.565,2367.785)
\blacken\path(669.000,1737.000)(729.000,1497.000)(789.000,1737.000)(729.000,1665.000)(669.000,1737.000)

\path(1927,1927)(1927,52)
\path(1927,1927)(3356,652)
\path(729,2389)(729,1497)

\path(2327,2327)(3462,1312)
\path(2276,2270)(2378,2384)
\path(3411,1255)(3513,1369)
\put(2874,1857){\makebox(0,0)[lb]{$R$}}

\put(4374,1857){\makebox(0,0)[lb]{$\Omega$}}
\put(3034.909,1913.773){\arc{2348.724}{5.7218}{6.7946}}

\put(2080,1335){\makebox(0,0)[lb]{$\varphi$}}
\put(2114.500,1428.250){\arc{676.041}{6.2277}{8.4420}}

\put(950,1857){\makebox(0,0)[lb]{$\vec{g}$}}

\put(3900,3400){\makebox(0,0)[lb]{\bf (a)}}

\end{picture} \hfill \hspace{-0.7cm}% GNUPLOT: LaTeX picture using EEPIC macros
\setlength{\unitlength}{0.225pt}
\begin{picture}(1049,900)(0,50)
\small
\thicklines \path(177,179)(197,179)
\thicklines \path(952,179)(932,179)
\put(155,179){\makebox(0,0)[r]{0}}
\thicklines \path(177,307)(197,307)
\thicklines \path(952,307)(932,307)
\put(155,307){\makebox(0,0)[r]{}}
\thicklines \path(177,434)(197,434)
\thicklines \path(952,434)(932,434)
\put(155,434){\makebox(0,0)[r]{10}}
\thicklines \path(177,562)(197,562)
\thicklines \path(952,562)(932,562)
\put(155,562){\makebox(0,0)[r]{}}
\thicklines \path(177,690)(197,690)
\thicklines \path(952,690)(932,690)
\put(155,690){\makebox(0,0)[r]{20}}
\thicklines \path(177,817)(197,817)
\thicklines \path(952,817)(932,817)
\put(155,817){\makebox(0,0)[r]{}}
\thicklines \path(177,945)(197,945)
\thicklines \path(952,945)(932,945)
\put(155,945){\makebox(0,0)[r]{30}}
\thicklines \path(177,179)(177,199)
\thicklines \path(177,945)(177,925)
\put(177,134){\makebox(0,0){0}}
\thicklines \path(306,179)(306,199)
\thicklines \path(306,945)(306,925)
\put(306,134){\makebox(0,0){}}
\thicklines \path(435,179)(435,199)
\thicklines \path(435,945)(435,925)
\put(435,134){\makebox(0,0){1}}
\thicklines \path(565,179)(565,199)
\thicklines \path(565,945)(565,925)
\put(565,134){\makebox(0,0){}}
\thicklines \path(694,179)(694,199)
\thicklines \path(694,945)(694,925)
\put(694,134){\makebox(0,0){2}}
\thicklines \path(823,179)(823,199)
\thicklines \path(823,945)(823,925)
\put(823,134){\makebox(0,0){}}
\thicklines \path(952,179)(952,199)
\thicklines \path(952,945)(952,925)
\put(952,134){\makebox(0,0){3}}
\thicklines \path(177,179)(952,179)(952,945)(177,945)(177,179)
\put(45,562){\makebox(0,0)[l]{\shortstack{$\varphi$ ($^\circ$)}}}
\put(564,50){\makebox(0,0){$t$ (s)}}
\put(875,562){\makebox(0,0){\bf (b)}}
\put(775,371){\makebox(0,0)[r]{simulation}}
\thinlines \path(797,371)(905,371)
\thinlines \path(179,186)(179,186)(182,202)(184,219)(187,236)(189,253)(191,270)(194,287)(196,304)(199,321)(201,337)(203,354)(206,371)(208,387)(211,404)(213,420)(215,437)(218,453)(220,469)(223,486)(225,502)(227,518)(230,535)(232,550)(235,567)(237,584)(239,601)(242,617)(244,633)(247,645)(249,662)(251,678)(254,694)(256,711)(259,726)(261,740)(263,756)(266,773)(268,789)(270,802)(273,819)(275,835)(278,849)(280,862)(282,878)(285,888)(287,900)(290,909)(292,914)(294,919)(297,923)
\thinlines \path(297,923)(299,923)(302,919)(304,914)(306,906)(309,897)(311,883)(314,871)(316,857)(318,844)(321,828)(323,811)(326,797)(328,781)(330,769)(333,756)(335,747)(338,739)(340,731)(342,726)(345,726)(347,726)(350,728)(352,735)(354,740)(357,752)(359,761)(362,773)(364,785)(366,802)(369,819)(371,831)(374,849)(376,862)(378,875)(381,888)(383,900)(386,909)(388,914)(390,919)(393,923)(395,923)(398,919)(400,914)(402,906)(405,897)(407,888)(410,875)(412,857)(414,845)(417,828)
\thinlines \path(417,828)(419,814)(421,799)(424,785)(426,773)(429,759)(431,747)(433,740)(436,731)(438,726)(441,726)(443,726)(445,726)(448,731)(450,740)(453,747)(455,761)(457,773)(460,785)(462,802)(465,814)(467,831)(469,845)(472,862)(474,875)(477,888)(479,897)(481,906)(484,914)(486,919)(489,923)(491,923)(493,919)(496,914)(498,909)(501,897)(503,888)(505,875)(508,862)(510,845)(513,831)(515,814)(517,802)(520,785)(522,773)(525,761)(527,749)(529,740)(532,735)(534,726)(537,726)
\thinlines \path(537,726)(539,726)(541,726)(544,731)(546,740)(549,747)(551,759)(553,769)(556,785)(558,799)(560,814)(563,828)(565,845)(568,857)(570,875)(572,887)(575,897)(577,906)(580,914)(582,919)(584,923)(587,923)(589,919)(592,914)(594,909)(596,900)(599,888)(601,875)(604,862)(606,849)(608,831)(611,819)(613,802)(616,789)(618,773)(620,761)(623,752)(625,740)(628,735)(630,728)(632,726)(635,726)(637,726)(640,731)(642,739)(644,747)(647,756)(649,768)(652,781)(654,797)(656,811)
\thinlines \path(656,811)(659,828)(661,840)(664,857)(666,871)(668,883)(671,897)(673,906)(676,914)(678,919)(680,923)(683,923)(685,919)(688,914)(690,909)(692,900)(695,892)(697,880)(700,866)(702,849)(704,837)(707,819)(709,802)(712,790)(714,776)(716,764)(719,752)(721,744)(724,735)(726,731)(728,726)(731,726)(733,726)(735,731)(738,739)(740,747)(743,756)(745,768)(747,781)(750,794)(752,811)(755,823)(757,840)(759,857)(762,871)(764,883)(767,897)(769,906)(771,914)(774,917)(776,923)
\thinlines \path(776,923)(779,923)(781,919)(783,917)(786,909)(788,900)(791,892)(793,880)(795,866)(798,852)(800,837)(803,819)(805,806)(807,790)(810,776)(812,764)(815,752)(817,744)(819,735)(822,731)(824,726)(827,726)(829,726)(831,731)(834,735)(836,744)(839,756)(841,764)(843,781)(846,794)(848,807)(851,823)(853,840)(855,854)(858,866)(860,883)(863,892)(865,906)(867,911)(870,917)(872,923)(875,923)(877,923)(879,917)(882,911)(884,902)(886,892)(889,880)(891,866)(894,854)(896,837)
\put(775,320){\makebox(0,0)[r]{arctan($\mu_w$)}}
\thinlines
\dottedline{12}(797,320)(905,320)
\dottedline{12}(177,822)(952,822)
\end{picture} }
  \caption[]{\parbox[t]{0.85\textwidth}{(a) Sketch of the quasi one-dimensional
             setup and (b) evolution of the angle as function of time in the
             simulation for $\mu_w=0.47$.}}
  \label{fig: one_part1}
\end{figure}
\fi

Equation~(\ref{eq: one_part1}) can be simplified by the following
steps~\cite{kassner}: setting $\widetilde\varphi\equiv \varphi-\varphi_0$ one
gets
\beq
  R\ddot{\widetilde\varphi} = - g\, \sqrt{1 + \mu_w^2}\, \sin (\widetilde\varphi)
                          + \mu_w R {\dot{\widetilde\varphi}}^2 \ .
\eeq
This equation does not exhibit an explicit time dependence and it is favourable
to use the transformation $u(\widetilde\varphi) \equiv \dot{\widetilde\varphi} 
\Rightarrow \ddot{\widetilde\varphi} = u' \dot{\widetilde\varphi} = u' u$ 
leading to the equation
\beq
  R u' u = - g\, \sqrt{1 + \mu_w^2}\, \sin (\widetilde\varphi) + \mu_w R u^2 \ .
\eeq
By using the transformation $v \equiv \frac{1}{2} u^2$ the equation can be
integrated and one obtains
\beq
  e^{-2\mu_w\widetilde\varphi} v(\widetilde\varphi) = - \frac{g\, 
      \sqrt{1 + \mu_w^2}}{R} \int_{-\varphi_0}^{\widetilde\varphi} e^{-2\mu_w\Psi} 
      \sin\Psi\, \textrm{d}\Psi + e^{2\mu_w\varphi}\, v(-\varphi_0) \ .
\eeq
Please note that the second term on the r.h.s.\ vanishes for
$\dot{\widetilde\varphi}(t=0)= 0$.

Backtransformation yields the following final equation for $\dot\varphi$
\beq
  {\dot{\varphi}}^2(t) = \frac{2 g}{R(1 + 4 \mu_w^2)} \left[ 3 \mu_w \sin \varphi +
                       (1 - 2 \mu_w^2) \left( \cos \varphi - e^{2 \mu_w \varphi}
		       \right) \right] + e^{2\mu_w\phi}\, \dot{\varphi}^2(0) \ .
  \label{eq: one_part2}
\eeq
It can be integrated numerically by separation of variable but the general
class of the motion can already be seen by a stability analysis. By setting
$\dot{\varphi}=\ddot{\varphi}=0$ in the original equation~(\ref{eq: one_part1}) 
one sees that the fixpoint must fulfill the condition
\[ \frac{\sin\varphi}{\cos\varphi} = \tan\varphi = \mu_w \]
which is nothing but the initial definition for $\varphi_0$. To study the
stability of the obtained fixpoint, one sets $\varphi(t) = \varphi_0
+ \varepsilon(t)$ in Eq.~(\ref{eq: one_part1}) and neglects higher terms in
$\varepsilon$ and $\dot{\varepsilon}$. The equation of motion for $\varepsilon$ 
reads
\[ \ddot{\varepsilon} = - \frac{g}{R\cos\varphi_0}\,\varepsilon \ . \]
This states that the fixpoint is marginally stable and the general motion is an
oscillation around it of the form
\beq
  \varphi(t) = \varphi_0 + A\, e^{i\omega t} + B\, e^{-i\omega t}
  \label{eq: focus}
\eeq
where
\[ \omega := \sqrt{\frac{g}{R\cos\varphi_0}} \ . \]
This oscillation can be seen in the simulation result shown in Fig.~\ref{fig:
one_part1}b. It is also found for the motion of a particle on a
plate~\cite{elmer97}. Please note that the frequency of the oscillation in
Eq.~(\ref{eq: focus}) does not depend on the external angular velocity,
$\Omega$, as long as the full-sliding condition is fulfilled.

\section{Numerical Model}
\label{sec: numeric}
In order to study the full dynamics of particles in a rotating drum, we use
two-dimensional molecular dynamics (MD) simulations. Particles are approximated
by spheres and only contact forces during collisions are
considered~\cite{ristow94b}. We use a {\em linear spring-dashpot model} for the
force in normal direction, reading
\beq F_n = - k_n h - \gamma_n \dot{h} \label{eq: fn}\eeq
and a viscous shear force in tangential direction, reading
\beq F_s = -\textrm{sign}(\gamma_s v_{\textrm{rel}})
         \min(\gamma_s|v_{\textrm{rel}}|, \mu|F_n|) \ . \label{eq: fs}\eeq
Here $h$ stands for the virtual overlap of the particle with the wall or with
another particle, $\mu$ denotes the friction coefficient and $v_{\textrm{rel}}$
the relative velocity of the two surfaces. The Coulomb friction law states
that the magnitude of the shear force can at most have a value equal to the
magnitude of the normal force multiplied by the friction coefficient. The model
parameter $k_n$ is related to the material stiffness, $\gamma_n$ to the normal
restitution coefficient, $e_n$, and $\gamma_s$ is chosen high enough  to ensure
that mostly the Coulomb friction law applies.

Using MD simulations, the angle $\varphi$ as function of time for a typical run
is shown in Fig.~\ref{fig: one_part1}b using the parameters $\Omega=1$~Hz,
$e_n=0.9$ and a particle-wall friction coefficient of $\mu_w=0.47$ with the 
initial condition $\varphi(0)=\dot{\varphi}(0)=0$. The oscillatory motion given 
by the analytic result, Eq.~(\ref{eq: focus}), can be seen as well. 

\section{Results and Discussion}
\label{sec: results}
In order to determine the maximum amplitude of this
motion, $\Delta\varphi$, one calculates the second turning point of the motion
by setting $\dot{\varphi}=0$ in Eq.~(\ref{eq: one_part2}) and using the initial
conditions $\varphi(0)=0$  and $\dot{\varphi}(0)=0$. One obtains
\beq 
  \sin\Delta\varphi = \frac{1-2\mu_w^2}{3\mu_w}\left( e^{2\mu_w\Delta\varphi} - 
                      \cos\Delta\varphi\right) \ , \ \Delta\varphi > 0 \ .
  \label{eq: one_part3}
\eeq
The sin-function on the l.h.s. can take only values from -1 to +1 and the
expression in parentheses on the r.h.s. is always greater than zero which gives
for small values of $\mu_w$ a solution close to zero. The fraction on the
r.h.s. changes sign at $\mu_w=1/\sqrt{2}\approx 0.7071$ leading to a diverging
solution around this value, e.g.\ for $\mu_w=0.71$, Eq.~(\ref{eq: one_part3})
does not have a solution. For a more quantitative analysis, the equation was
integrated numerically using {\sc Mathematica} and the obtained solution is
shown as solid line in Fig.~\ref{fig: one_part2}a as function of the friction
coefficient $\mu_w$. Even though solutions of $\Delta\varphi > 90^\circ$ are
possible analytically from Eq.~(\ref{eq: one_part3}), they do not lead to an 
oscillating motion since for $\varphi > 90^\circ$ the particle will detach from
the wall on its way downwards which invalidates the assumptions leading to
Eq.~(\ref{eq: one_part1}). We added in Fig.~\ref{fig: one_part2}a the data for
$\Delta\varphi$ from the MD simulations as open circles ($\circ$). These were
obtained by measuring the angle difference via the maxima and minima in plots
like the one shown in Fig.~\ref{fig: one_part1}b. The values vary very little in
time and the agreement with the analytic expression, taken from Eq.~(\ref{eq:
one_part3}), is perfect, thus demonstrating that the simulation indeed describes
the full-sliding regime when a viscous shear force is used together with the
Coulomb threshold criterion, Eq.~(\ref{eq: fs}). For angles $\varphi$ greater
than 90 degrees, i.e.\ points lying above the dotted line in Fig.~\ref{fig:
one_part2}a, the value of $\Delta\varphi$ was obtained by taking the maximum
value during the first oscillation just before the free fall during the
downwards motion sets in. 
\ifx\draft\undefined
\else
\begin{figure}[t]
  \hbox{\hspace{-0.7cm}% GNUPLOT: LaTeX picture using EEPIC macros
\setlength{\unitlength}{0.225pt}
\begin{picture}(1049,900)(0,50)
\small
\thicklines \path(199,179)(219,179)
\thicklines \path(952,179)(932,179)
\put(177,179){\makebox(0,0)[r]{0}}
\thicklines \path(199,300)(219,300)
\thicklines \path(952,300)(932,300)
%\put(177,300){\makebox(0,0)[r]{30}}
\thicklines \path(199,421)(219,421)
\thicklines \path(952,421)(932,421)
\put(177,421){\makebox(0,0)[r]{60}}
\thicklines \path(199,542)(219,542)
\thicklines \path(952,542)(932,542)
%\put(177,542){\makebox(0,0)[r]{90}}
\thicklines \path(199,663)(219,663)
\thicklines \path(952,663)(932,663)
\put(177,663){\makebox(0,0)[r]{120}}
\thicklines \path(199,784)(219,784)
\thicklines \path(952,784)(932,784)
%\put(177,784){\makebox(0,0)[r]{150}}
\thicklines \path(199,905)(219,905)
\thicklines \path(952,905)(932,905)
\put(177,905){\makebox(0,0)[r]{180}}
\thicklines \path(199,179)(199,199)
\thicklines \path(199,945)(199,925)
\put(199,134){\makebox(0,0){0}}
\thicklines \path(299,179)(299,199)
\thicklines \path(299,945)(299,925)
%\put(299,134){\makebox(0,0){0.1}}
\thicklines \path(400,179)(400,199)
\thicklines \path(400,945)(400,925)
\put(400,134){\makebox(0,0){0.2}}
\thicklines \path(500,179)(500,199)
\thicklines \path(500,945)(500,925)
%\put(500,134){\makebox(0,0){0.3}}
\thicklines \path(601,179)(601,199)
\thicklines \path(601,945)(601,925)
\put(601,134){\makebox(0,0){0.4}}
\thicklines \path(701,179)(701,199)
\thicklines \path(701,945)(701,925)
%\put(701,134){\makebox(0,0){0.5}}
\thicklines \path(801,179)(801,199)
\thicklines \path(801,945)(801,925)
\put(801,134){\makebox(0,0){0.6}}
\thicklines \path(902,179)(902,199)
\thicklines \path(902,945)(902,925)
%\put(902,134){\makebox(0,0){0.7}}
\thicklines \path(199,179)(952,179)(952,945)(199,945)(199,179)
\put(25,542){\makebox(0,0)[l]{\shortstack{$\Delta \varphi$ ($^\circ$)}}}
\put(575,50){\makebox(0,0){$\mu_w$}}
\put(877,421){\makebox(0,0){\bf (a)}}
\put(478,824){\makebox(0,0)[r]{analytic}}
\thinlines \path(500,824)(608,824)
\thinlines \path(199,179)(199,179)(223,190)(246,201)(270,212)(294,223)(317,234)(341,245)(365,256)(389,267)(412,279)(436,290)(460,302)(483,314)(507,326)(531,339)(554,351)(578,365)(602,378)(625,393)(649,408)(673,424)(697,440)(720,459)(744,479)(768,501)(791,526)(815,555)(839,591)(862,639)(886,711)(910,933)
\put(478,779){\makebox(0,0)[r]{simulation}}
\put(249,203){\circle{18}}
\put(299,226){\circle{18}}
\put(350,250){\circle{18}}
\put(400,274){\circle{18}}
\put(450,299){\circle{18}}
\put(500,325){\circle{18}}
\put(550,351){\circle{18}}
\put(601,380){\circle{18}}
\put(651,410){\circle{18}}
\put(701,446){\circle{18}}
\put(751,486){\circle{18}}
\put(801,541){\circle{18}}
\put(852,618){\circle{18}}
\put(872,671){\circle{18}}
\put(882,711){\circle{18}}
\put(892,778){\circle{18}}
\put(554,779){\circle{18}}
\thinlines
\dottedline{12}(199,542)(952,542)
\end{picture} \hfill \hspace{-0.7cm}% GNUPLOT: LaTeX picture using EEPIC macros
\setlength{\unitlength}{0.225pt}
\begin{picture}(1049,900)(0,50)
\small
\thicklines \path(177,179)(197,179)
\thicklines \path(952,179)(932,179)
\put(155,179){\makebox(0,0)[r]{15}}
\thicklines \path(177,371)(197,371)
\thicklines \path(952,371)(932,371)
\put(155,371){\makebox(0,0)[r]{20}}
\thicklines \path(177,562)(197,562)
\thicklines \path(952,562)(932,562)
\put(155,562){\makebox(0,0)[r]{25}}
\thicklines \path(177,754)(197,754)
\thicklines \path(952,754)(932,754)
\put(155,754){\makebox(0,0)[r]{30}}
\thicklines \path(177,945)(197,945)
\thicklines \path(952,945)(932,945)
\put(155,945){\makebox(0,0)[r]{35}}
\thicklines \path(177,179)(177,199)
\thicklines \path(177,945)(177,925)
\put(177,134){\makebox(0,0){0}}
\thicklines \path(371,179)(371,199)
\thicklines \path(371,945)(371,925)
\put(371,134){\makebox(0,0){5}}
\thicklines \path(565,179)(565,199)
\thicklines \path(565,945)(565,925)
\put(565,134){\makebox(0,0){10}}
\thicklines \path(758,179)(758,199)
\thicklines \path(758,945)(758,925)
\put(758,134){\makebox(0,0){15}}
\thicklines \path(952,179)(952,199)
\thicklines \path(952,945)(952,925)
\put(952,134){\makebox(0,0){20}}
\thicklines \path(177,179)(952,179)(952,945)(177,945)(177,179)
\put(25,658){\makebox(0,0)[l]{\shortstack{$\langle \varphi \rangle$ ($^\circ$)}}}
\put(564,50){\makebox(0,0){$\Omega$ (Hz)}}
\put(875,562){\makebox(0,0){\bf (b)}}
\thinlines
\put(775,371){\makebox(0,0)[r]{experiment}}
\dottedline{12}(797,371)(905,371)
\put(196,719){\circle*{12}}
\put(197,731){\circle*{12}}
\put(198,685){\circle*{12}}
\put(200,696){\circle*{12}}
\put(200,684){\circle*{12}}
\put(201,686){\circle*{12}}
\put(203,399){\circle*{12}}
\put(204,374){\circle*{12}}
\put(207,243){\circle*{12}}
\put(211,317){\circle*{12}}
\put(214,408){\circle*{12}}
\put(221,584){\circle*{12}}
\put(226,569){\circle*{12}}
\put(231,568){\circle*{12}}
\put(235,578){\circle*{12}}
\put(242,536){\circle*{12}}
\put(247,554){\circle*{12}}
\put(258,616){\circle*{12}}
\put(267,633){\circle*{12}}
\put(271,644){\circle*{12}}
\put(305,615){\circle*{12}}
\put(311,615){\circle*{12}}
\put(324,642){\circle*{12}}
\put(333,784){\circle*{12}}
\put(344,672){\circle*{12}}
\put(357,665){\circle*{12}}
\put(375,796){\circle*{12}}
\put(389,754){\circle*{12}}
\put(402,735){\circle*{12}}
\put(479,765){\circle*{12}}
\put(503,690){\circle*{12}}
\put(550,777){\circle*{12}}
\put(907,802){\circle*{12}}
\put(851,371){\circle*{12}}
\dottedline{12}(196,719)(197,731)(198,685)(200,696)(200,684)(201,686)(203,399)
\dottedline{12}(203,399)(204,374)(207,243)(211,317)(214,408)(221,584)(226,569)
\dottedline{12}(226,569)(231,568)(235,578)(242,536)(247,554)(258,616)(267,633)
\dottedline{12}(267,633)(271,644)(305,615)(311,615)(324,642)(333,784)(344,672)
\dottedline{12}(344,672)(357,665)(375,796)(389,754)(402,735)(479,765)(503,690)
\dottedline{12}(503,690)(550,777)(907,802)

\thicklines
\put(775,326){\makebox(0,0)[r]{simulation}}
\thinlines \path(797,326)(905,326)
\thinlines \path(196,568)(196,568)(216,572)(255,583)(293,600)(332,629)(371,663)(390,681)(410,704)(429,729)(448,754)(468,758)(487,758)(565,758)(661,758)(758,758)(855,763)
\put(196,568){\circle{18}}
\put(216,572){\circle{18}}
\put(255,583){\circle{18}}
\put(293,600){\circle{18}}
\put(332,629){\circle{18}}
\put(371,663){\circle{18}}
\put(390,681){\circle{18}}
\put(410,704){\circle{18}}
\put(429,729){\circle{18}}
\put(448,754){\circle{18}}
\put(468,758){\circle{18}}
\put(487,758){\circle{18}}
\put(565,758){\circle{18}}
\put(661,758){\circle{18}}
\put(758,758){\circle{18}}
\put(855,763){\circle{18}}
\put(851,326){\circle{18}}
\end{picture} }
  \caption[]{\parbox[t]{0.85\textwidth}{(a) Comparison of the analytic
             expression for the maximum amplitude with the simulation and (b) 
             frequency dependence of the average angle for the experiment 
             ($\bullet$) and the simulation using $\mu_w=0.47$ ($\circ$).}}
  \label{fig: one_part2}
\end{figure}
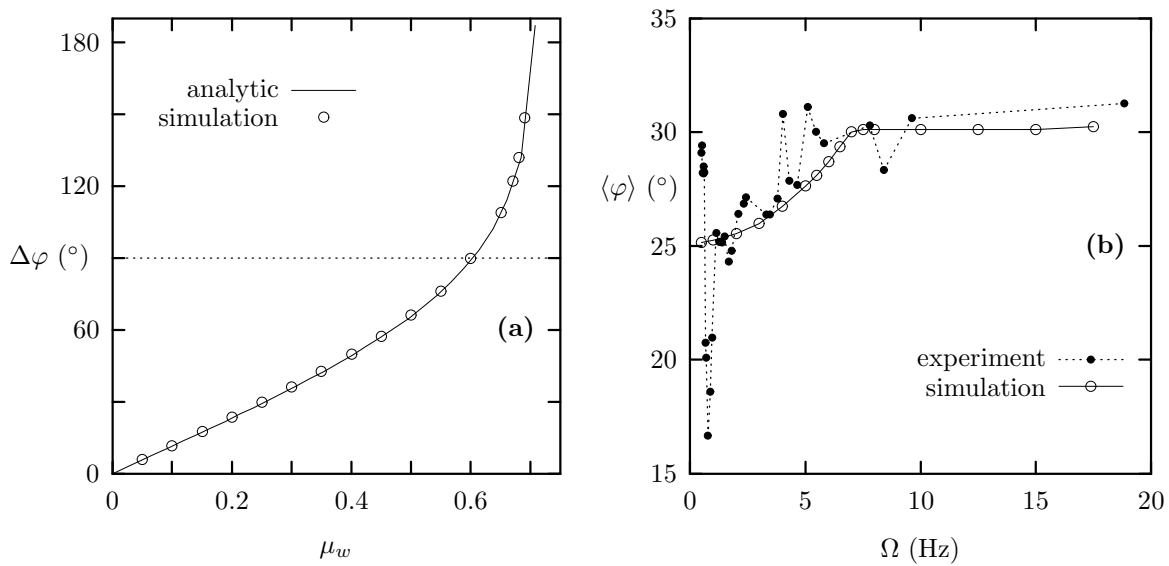
\fi

The maximum amplitude is only reached for a high enough value of the rotation
speed of the drum, $\Omega$, for the chosen initial conditions. For lower
values of $\Omega$, the system will oscillate with an intermediate amplitude
since they are all marginally stable. If the initial conditions are chosen such
that the initial amplitude is greater than $\Delta\varphi$, the system will
dissipate energy and finally reach a state with an amplitude lower than or
equal to $\Delta\varphi$. In the numerical simulations, the frequency of the
oscillation shows no dependence on the rotation speed $\Omega$ in agreement
with the analytic result, Eq.~(\ref{eq: focus}), as long as the full-sliding
condition is fulfilled. The frequency is 16.5 Hz for the simulation, extracted
from Fig.~\ref{fig: one_part1}b, and 16.9 Hz for the analytic expression,
Eq.~(\ref{eq: focus}). A similar system was investigated
experimentally~\cite{betat97} by gluing two spheres together and slightly
tilting the drum in order to force the particle to slide along one of the end
caps. There, only a weak frequency dependence was found for values of $\Omega >
2.5$~Hz giving a dominant oscillation frequency of 15.7 Hz, but for lower
values, a stick-slip motion was clearly visible and led to lower values of the
dominant frequency mode in the oscillation. Since no sticking was considered in
the analytic model, it cannot capture this feature. The numerical model assumes
a perfectly shaped drum whereas in the experiment a slight aspherity was
observed and additional friction with one end cap was present thus amplifying
the stick-slip region. Keeping these differences in mind the agreement is very 
satisfactory. By looking at the mean angle, $\langle\varphi\rangle$, of the
particle as function of $\Omega$, a similar frequency dependence was found in
the experiment. To a certain extent, this is also visible in the MD simulations
and we show in Fig.~\ref{fig:  one_part2}b the experimental points, taken
from~\cite{betat97}, as filled circles ($\bullet$) and the numerical data by
open circles ($\circ$). For $\Omega \lesssim 7$ Hz, the mean angle decreases
with decreasing angular frequency of the drum and the general trend is well
captured by the simulation. The large fluctuations in the experimental data can
probably be attributed to the stick-slip motion which was enhanced by a slight
aspherity of the drum. The best agreement with the experimental data in the
high angular frequency regime was obtained for $\mu_w=0.47$.
\ifx\draft\undefined
\else
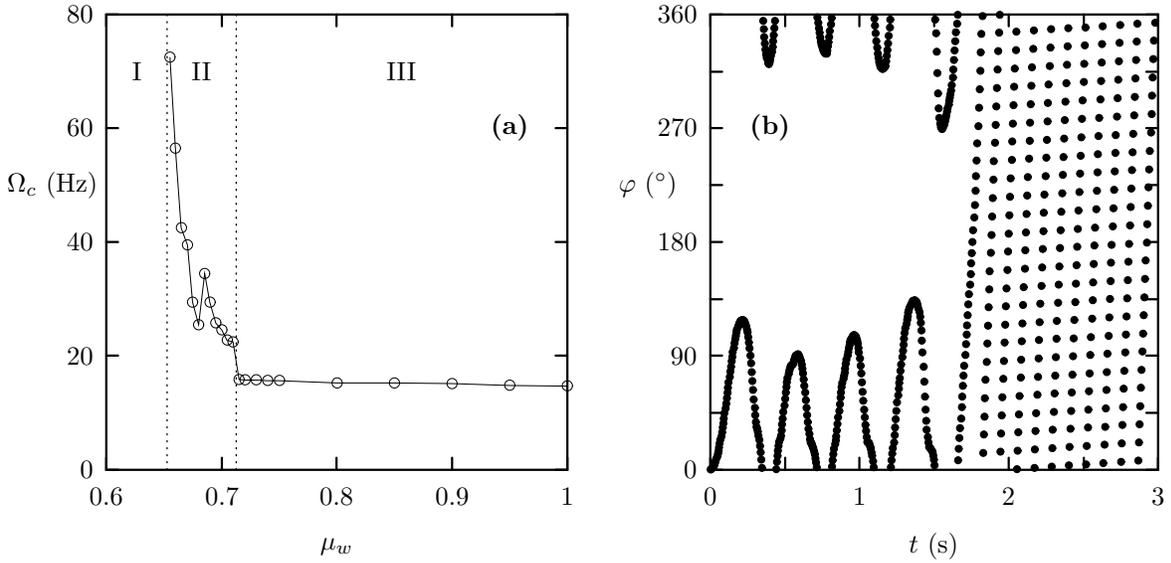
\begin{figure}[t]
  \hbox{\hspace{-0.7cm}% GNUPLOT: LaTeX picture using EEPIC macros
\setlength{\unitlength}{0.225pt}
\begin{picture}(1049,900)(0,50)
\small
\thicklines \path(177,179)(197,179)
\thicklines \path(952,179)(932,179)
\put(155,179){\makebox(0,0)[r]{0}}
\thicklines \path(177,371)(197,371)
\thicklines \path(952,371)(932,371)
\put(155,371){\makebox(0,0)[r]{20}}
\thicklines \path(177,562)(197,562)
\thicklines \path(952,562)(932,562)
\put(155,562){\makebox(0,0)[r]{40}}
\thicklines \path(177,754)(197,754)
\thicklines \path(952,754)(932,754)
\put(155,754){\makebox(0,0)[r]{60}}
\thicklines \path(177,945)(197,945)
\thicklines \path(952,945)(932,945)
\put(155,945){\makebox(0,0)[r]{80}}
\thicklines \path(177,179)(177,199)
\thicklines \path(177,945)(177,925)
\put(177,134){\makebox(0,0){0.6}}
\thicklines \path(371,179)(371,199)
\thicklines \path(371,945)(371,925)
\put(371,134){\makebox(0,0){0.7}}
\thicklines \path(564,179)(564,199)
\thicklines \path(564,945)(564,925)
\put(564,134){\makebox(0,0){0.8}}
\thicklines \path(758,179)(758,199)
\thicklines \path(758,945)(758,925)
\put(758,134){\makebox(0,0){0.9}}
\thicklines \path(952,179)(952,199)
\thicklines \path(952,945)(952,925)
\put(952,134){\makebox(0,0){1}}
\thicklines \path(177,179)(952,179)(952,945)(177,945)(177,179)
\put(10,658){\makebox(0,0)[l]{\shortstack{$\Omega_c$ (Hz)}}}
\put(564,50){\makebox(0,0){$\mu_w$}}
\put(228,849){\makebox(0,0){I}}
\put(337,849){\makebox(0,0){II}}
\put(673,849){\makebox(0,0){III}}
\put(855,754){\makebox(0,0){\bf (a)}}
\thinlines
\path(952,320)(952,320)(855,321)(758,324)(661,325)(565,325)(468,329)(448,329)(429,330)(410,330)(400,331)(390,394)(380,397)(371,414)(361,426)(351,461)(342,509)(332,423)(322,461)(313,557)(303,586)(293,720)(284,873)
\put(952,320){\circle{18}}
\put(855,321){\circle{18}}
\put(758,324){\circle{18}}
\put(661,325){\circle{18}}
\put(565,325){\circle{18}}
\put(468,329){\circle{18}}
\put(448,329){\circle{18}}
\put(429,330){\circle{18}}
\put(410,330){\circle{18}}
\put(400,331){\circle{18}}
\put(390,394){\circle{18}}
\put(380,397){\circle{18}}
\put(371,414){\circle{18}}
\put(361,426){\circle{18}}
\put(351,461){\circle{18}}
\put(342,509){\circle{18}}
\put(332,423){\circle{18}}
\put(322,461){\circle{18}}
\put(313,557){\circle{18}}
\put(303,586){\circle{18}}
\put(293,720){\circle{18}}
\put(284,873){\circle{18}}
%\put(279,179){\circle{18}}
%\put(279,945){\circle{18}}
%\put(395,179){\circle{18}}
%\put(395,945){\circle{18}}
\dottedline{12}(279,179)(279,945)
\dottedline{12}(395,179)(395,945)
\end{picture} \hfill \hspace{-0.7cm}\input few_f3b }
  \caption[]{\parbox[t]{0.85\textwidth}{(a) Transition frequency to the
             centrifugal regime as function of friction coefficient and (b)
             angle of the particle as function of time for $\mu_w=0.665$ and
             $\Omega=43$ Hz (regime II).}}
  \label{fig: one_part3}
\end{figure}
\fi

The solution of the analytic Eq.~(\ref{eq: one_part3}) and the corresponding
Fig.~\ref{fig: one_part2}a state that the amplitude of the oscillation will be
larger than 180$^\circ$ ($\stackrel{\wedge}{=} \pi$) for values of $\mu
\gtrsim  0.7$ for a sufficiently high value of $\Omega$. Physically that means
that the particle will make overturns following the external sense of rotation
and touches the wall all the time. This regime will be called the {\em
centrifugal regime} and marked by the symbol III. It exists a well defined
transition frequency which we will call $\Omega_c$. The question now arises
under what conditions this centrifugal regime exists and if it will be possible
to make immediate overturns starting with a particle at rest. To address this
question, we investigated numerically the transition frequency, $\Omega_c$, as
function of friction coefficient, $\mu$, for a particle initially at rest. It
can be viewed as a phase diagram and is shown in Fig.~\ref{fig: one_part3}a.
Three different regimes can be identified: (I) regardless of the angular
frequency and the initial conditions the final state will oscillate below an
angle of 90 degrees showing no overturns. (II) above a threshold of $\mu
\approx 0.65$, the centrifugal regime can be reached and the transition
frequency decreases with increasing value of $\mu$. Since the dynamics depends
on the initial state, fluctuations in $\Omega_c$ are visible. To illustrate
this point, we show in Fig.~\ref{fig: one_part3}b a typical trajectory,
particle angle $\varphi$ as function of time, in this regime for the parameters
$\mu=0.665$ and  $\Omega=43$~Hz. After undergoing oscillations with different
amplitudes initially, the particle reaches the centrifugal regime for
$t>1.7$~s. (III) for values of $\mu \gtrsim 0.7$ the particle will reach the
centrifugal regime within the first external rotation. The transition frequency
is independent of the friction coefficient in this regime.

\ifx\draft\undefined
\else
\begin{figure}[t]
  \hbox{% GNUPLOT: LaTeX picture using EEPIC macros
\setlength{\unitlength}{0.22pt}
\begin{picture}(960,900)(70,50)
%\tenrm
\thicklines \path(199,179)(219,179)
\thicklines \path(952,179)(932,179)
\put(177,179){\makebox(0,0)[r]{-15}}
\thicklines \path(199,307)(219,307)
\thicklines \path(952,307)(932,307)
\put(177,307){\makebox(0,0)[r]{-10}}
\thicklines \path(199,434)(219,434)
\thicklines \path(952,434)(932,434)
\put(177,434){\makebox(0,0)[r]{-5}}
\thicklines \path(199,562)(219,562)
\thicklines \path(952,562)(932,562)
\put(177,562){\makebox(0,0)[r]{0}}
\thicklines \path(199,690)(219,690)
\thicklines \path(952,690)(932,690)
\put(177,690){\makebox(0,0)[r]{5}}
\thicklines \path(199,817)(219,817)
\thicklines \path(952,817)(932,817)
\put(177,817){\makebox(0,0)[r]{10}}
\thicklines \path(199,945)(219,945)
\thicklines \path(952,945)(932,945)
\put(177,945){\makebox(0,0)[r]{15}}
\thicklines \path(199,179)(199,199)
\thicklines \path(199,945)(199,925)
\put(199,134){\makebox(0,0){0}}
\thicklines \path(387,179)(387,199)
\thicklines \path(387,945)(387,925)
\put(387,134){\makebox(0,0){$\pi/8$}}
\thicklines \path(576,179)(576,199)
\thicklines \path(576,945)(576,925)
\put(576,134){\makebox(0,0){$\pi/4$}}
\thicklines \path(764,179)(764,199)
\thicklines \path(764,945)(764,925)
\put(764,134){\makebox(0,0){$3\pi/8$}}
\thicklines \path(952,179)(952,199)
\thicklines \path(952,945)(952,925)
\put(952,134){\makebox(0,0){$\pi/2$}}
\thicklines \path(199,179)(952,179)(952,945)(199,945)(199,179)
\put(60,562){\makebox(0,0)[l]{\shortstack{$\dot{\varphi}$}}}
\put(575,40){\makebox(0,0){$\varphi$}}
\thicklines
\ifx\color\undefined\thinlines\else\color{red}\fi
%\put(778,903){\makebox(0,0)[r]{$\Omega$=1 Hz}}
%\path(800,903)(908,903)
\path(203,588)(203,588)(209,587)(215,587)(222,587)(228,587)(234,587)(240,587)(246,587)(252,587)(258,587)(265,587)(271,587)(277,587)(283,587)(289,587)(295,587)(301,587)(307,587)(313,587)(319,587)(325,587)(331,587)(337,587)(343,587)(349,587)(355,587)(361,587)(367,586)(373,586)(379,586)(385,586)(391,586)(396,586)(402,586)(408,586)(414,586)(420,586)(426,586)(432,586)(437,586)(443,586)(449,586)(455,586)(461,586)(467,585)(473,584)(478,581)(482,578)(485,575)(488,570)
\path(488,570)(489,566)(489,561)(488,557)(487,552)(484,548)(480,545)(476,542)(471,540)(465,538)(459,538)(453,538)(448,540)(442,542)(438,544)(434,548)(432,552)(429,556)(428,560)(428,564)(429,569)(432,573)(435,577)(438,580)(443,583)(448,584)(454,586)(460,586)(466,585)(471,584)(476,582)
\ifx\color\undefined\thicklines\else\color{green}\fi
%\put(778,858){\makebox(0,0)[r]{2 Hz}}
%\path(800,858)(908,858)
\path(203,598)(203,598)(215,613)(227,613)(240,613)(252,613)(264,613)(277,613)(289,613)(301,612)(314,612)(326,612)(338,612)(350,612)(362,612)(375,612)(386,612)(399,612)(411,612)(423,612)(435,612)(447,612)(459,612)(471,611)(483,609)(494,605)(504,600)(512,593)(519,584)(523,575)(525,566)(524,556)(522,546)(517,537)(510,529)(501,523)(491,518)(480,514)(468,513)(456,513)(444,515)(433,518)(423,523)(414,529)(407,537)(402,545)(399,553)(398,562)(399,571)(402,579)(407,587)
\path(407,587)(414,594)(423,601)(433,606)(444,609)(456,611)(468,611)(479,610)(491,606)(501,601)(510,595)
\ifx\color\undefined\Thicklines\else\color{blue}\fi
%\put(778,813){\makebox(0,0)[r]{5 Hz}}
%\path(800,813)(908,813)
\path(203,598)(203,598)(217,634)(238,669)(268,689)(299,689)(330,689)(360,689)(391,688)(422,688)(453,688)(483,688)(514,687)(544,681)(571,670)(596,655)(616,636)(632,613)(641,588)(643,561)(640,535)(631,510)(615,487)(595,468)(570,453)(542,443)(512,437)(482,436)(452,440)(422,447)(396,458)(372,472)(352,488)(336,506)(325,525)(319,545)(317,566)(320,586)(329,606)(341,625)(359,642)(381,657)(405,670)(432,680)(463,686)(493,688)(523,685)
\thicklines
%\ifx\color\undefined\thinlines\else\color{cyan}\fi
%\put(778,768){\makebox(0,0)[r]{11 Hz}}
%\path(800,768)(908,768)
%\ifx\color\undefined
\dottedline{12}(203,598)(203,598)(217,634)(238,669)(269,703)(307,736)(353,765)(406,791)(464,812)(525,816)(587,815)(648,811)(708,798)(762,775)(811,742)(849,701)(877,652)(893,598)(894,542)(883,488)(859,437)(823,393)(777,358)(724,332)(667,316)(606,309)(544,312)(484,322)(427,339)(376,361)(330,387)(291,416)(259,448)(236,481)(220,515)(213,549)(214,584)(223,618)(241,652)(267,685)(300,716)(341,744)(389,770)(442,791)(499,806)(560,814)(622,814)
%\else
%\path(203,598)(203,598)(217,634)(238,669)(269,703)(307,736)(353,765)(406,791)(464,812)(525,816)(587,815)(648,811)(708,798)(762,775)(811,742)(849,701)(877,652)(893,598)(894,542)(883,488)(859,437)(823,393)(777,358)(724,332)(667,316)(606,309)(544,312)(484,322)(427,339)(376,361)(330,387)(291,416)(259,448)(236,481)(220,515)(213,549)(214,584)(223,618)(241,652)(267,685)(300,716)(341,744)(389,770)(442,791)(499,806)(560,814)(622,814)
%\fi
%\ifx\color\undefined\thinlines\else\color{magenta}\fi
%\put(778,723){\makebox(0,0)[r]{20 Hz}}
%\path(800,723)(908,723)
%\ifx\color\undefined
\drawline[-50](201,584)(205,605)(213,627)(224,648)(238,669)(256,690)(276,710)(299,729)(325,748)(353,765)(384,781)(417,796)(452,808)(488,820)(527,828)(565,833)(606,835)(645,833)(684,830)(723,822)(760,810)(795,794)(827,774)(857,751)(883,725)(904,696)(921,664)(934,630)(941,595)(944,559)(940,523)(932,488)(919,455)(901,423)(878,394)(852,368)(822,346)(790,327)(754,312)(717,302)(678,294)(638,289)(599,289)(559,291)(520,299)(483,307)(446,318)(412,330)(378,345)(348,362)(320,379)(294,398)(272,417)(253,437)(236,458)(222,479)(212,500)(204,522)(200,544)(199,566)
\end{picture} \hfill % GNUPLOT: LaTeX picture using EEPIC macros
\setlength{\unitlength}{0.22pt}
\begin{picture}(960,900)(60,50)
%\tenrm
\thicklines \path(221,270)(241,270)
\thicklines \path(952,270)(932,270)
\put(199,270){\makebox(0,0)[r]{}}
\thicklines \path(221,366)(241,366)
\thicklines \path(952,366)(932,366)
\put(199,366){\makebox(0,0)[r]{-0.5}}
\thicklines \path(221,463)(241,463)
\thicklines \path(952,463)(932,463)
\put(199,463){\makebox(0,0)[r]{}}
\thicklines \path(221,559)(241,559)
\thicklines \path(952,559)(932,559)
\put(199,559){\makebox(0,0)[r]{0}}
\thicklines \path(221,656)(241,656)
\thicklines \path(952,656)(932,656)
\put(199,656){\makebox(0,0)[r]{}}
\thicklines \path(221,752)(241,752)
\thicklines \path(952,752)(932,752)
\put(199,752){\makebox(0,0)[r]{0.5}}
\thicklines \path(221,849)(241,849)
\thicklines \path(952,849)(932,849)
\put(199,849){\makebox(0,0)[r]{}}
\thicklines \path(221,945)(241,945)
\thicklines \path(952,945)(932,945)
\put(199,945){\makebox(0,0)[r]{1}}
\thicklines \path(221,179)(221,199)
\thicklines \path(221,945)(221,925)
\put(221,134){\makebox(0,0){0}}
\thicklines \path(404,179)(404,199)
\thicklines \path(404,945)(404,925)
\put(404,134){\makebox(0,0){$\pi/8$}}
\thicklines \path(587,179)(587,199)
\thicklines \path(587,945)(587,925)
\put(587,134){\makebox(0,0){$\pi/4$}}
\thicklines \path(769,179)(769,199)
\thicklines \path(769,945)(769,925)
\put(769,134){\makebox(0,0){$3\pi/8$}}
\thicklines \path(952,179)(952,199)
\thicklines \path(952,945)(952,925)
\put(952,134){\makebox(0,0){$\pi/2$}}
\thicklines \path(221,179)(952,179)(952,945)(221,945)(221,179)
\put(80,562){\makebox(0,0)[l]{\shortstack{\large $\frac{\dot{\varphi}}{\Omega}$}}}
\put(586,40){\makebox(0,0){$\varphi$}}
\thicklines
\ifx\color\undefined\thinlines\else\color{red}\fi
%\put(778,903){\makebox(0,0)[r]{$\Omega$=1 Hz}}
%\path(800,903)(908,903)
\path(225,945)(225,945)(231,945)(237,944)(243,943)(249,943)(255,942)(261,942)(267,941)(273,940)(279,940)(285,939)(291,938)(296,938)(302,937)(308,937)(314,936)(320,936)(326,935)(331,935)(338,934)(344,933)(349,933)(355,932)(361,932)(366,931)(373,931)(379,930)(384,930)(390,929)(396,928)(401,928)(407,928)(413,927)(418,926)(424,926)(430,925)(435,925)(441,925)(447,924)(452,923)(458,923)(464,923)(470,922)(475,921)(481,910)(487,887)(491,851)(496,804)(499,749)(501,686)
\path(501,686)(503,618)(503,548)(502,479)(500,412)(498,351)(494,298)(490,255)(485,223)(479,204)(474,197)(468,203)(463,222)(457,253)(453,294)(449,343)(447,401)(444,463)(444,529)(444,596)(444,661)(447,723)(450,780)(453,829)(458,869)(463,899)(469,916)(474,921)(480,914)(485,893)(490,860)
\ifx\color\undefined\thicklines\else\color{green}\fi
%\put(778,858){\makebox(0,0)[r]{2 Hz}}
%\path(800,858)(908,858)
\path(225,833)(225,833)(237,945)(249,944)(261,943)(273,943)(285,942)(296,942)(309,941)(320,940)(332,940)(344,939)(356,939)(368,938)(379,938)(392,937)(403,937)(415,936)(426,936)(439,935)(450,935)(462,934)(474,933)(485,930)(496,914)(507,886)(517,844)(525,791)(531,729)(535,660)(538,587)(537,513)(535,440)(530,373)(523,313)(514,263)(504,225)(494,199)(483,186)(470,187)(459,202)(448,228)(439,265)(430,312)(423,367)(418,427)(415,492)(414,558)(415,625)(418,689)(423,750)
\path(423,750)(430,805)(439,852)(448,889)(459,916)(470,931)(482,932)(493,920)(504,895)(514,857)(523,807)
\ifx\color\undefined\Thicklines\else\color{blue}\fi
%\put(778,813){\makebox(0,0)[r]{5 Hz}}
%\path(800,813)(908,813)
\path(225,669)(225,669)(238,777)(259,884)(288,944)(318,943)(348,943)(378,942)(408,942)(438,941)(467,941)(497,940)(526,936)(556,918)(582,886)(606,840)(626,782)(641,713)(650,637)(652,557)(649,477)(640,401)(625,333)(605,275)(581,230)(554,199)(525,182)(496,179)(466,189)(438,212)(412,245)(389,287)(370,335)(354,390)(344,448)(337,509)(336,571)(339,632)(347,692)(359,749)(376,802)(397,848)(421,887)(448,916)(477,934)(506,940)(535,932)
\thicklines
%\ifx\color\undefined\thinlines\else\color{cyan}\fi
%\put(778,768){\makebox(0,0)[r]{11 Hz}}
%\path(800,768)(908,768)
%\ifx\color\undefined
\dottedline{12}(225,609)(225,609)(238,659)(259,707)(289,754)(326,798)(370,839)(422,874)(478,903)(539,924)(603,934)(667,931)(729,916)(789,886)(841,842)(885,784)(917,714)(937,637)(944,555)(936,474)(915,397)(881,329)(836,272)(783,229)(724,201)(660,186)(596,186)(533,197)(472,218)(416,248)(366,284)(322,325)(285,369)(257,416)(236,465)(224,515)(221,564)(226,614)(240,664)(262,712)(292,759)
%\else
%\path(225,609)(225,609)(238,659)(259,707)(289,754)(326,798)(370,839)(422,874)(478,903)(539,924)(603,934)(667,931)(729,916)(789,886)(841,842)(885,784)(917,714)(937,637)(944,555)(936,474)(915,397)(881,329)(836,272)(783,229)(724,201)(660,186)(596,186)(533,197)(472,218)(416,248)(366,284)(322,325)(285,369)(257,416)(236,465)(224,515)(221,564)(226,614)(240,664)(262,712)(292,759)
%\fi
%\ifx\color\undefined\thinlines\else\color{magenta}\fi
%\put(778,723){\makebox(0,0)[r]{20 Hz}}
%\path(800,723)(908,723)
%\ifx\color\undefined
\drawline[-50](223,576)(227,592)(235,608)(246,625)(259,640)(276,656)(295,671)(318,686)(343,700)(370,713)(401,725)(432,736)(466,746)(502,754)(539,760)(577,764)(616,765)(654,765)(692,762)(729,755)(766,747)(799,735)(831,720)(859,703)(885,682)(906,660)(922,636)(934,611)(941,584)(944,557)(941,530)(933,504)(920,478)(902,454)(881,433)(855,413)(826,396)(794,382)(760,371)(724,362)(686,356)(647,354)(609,353)(570,355)(533,360)(496,366)(461,374)(427,384)(395,396)(366,408)(339,421)(314,435)(292,450)(273,465)(257,481)(244,497)(233,513)(226,529)(222,546)(221,562)
\end{picture} }
  \caption[]{\parbox[t]{0.85\textwidth}{Phase space plot for different
             angular frequencies $\Omega\!=\!1($---$),2(${\bf ---}$),5 \\
             $({\bf ---})$,11(\cdot \cdot \cdot),20($-- --) Hz, (a) 
             unnormalized and (b) normalized by $\Omega$.}}
  \label{fig: one_part4}
\end{figure}
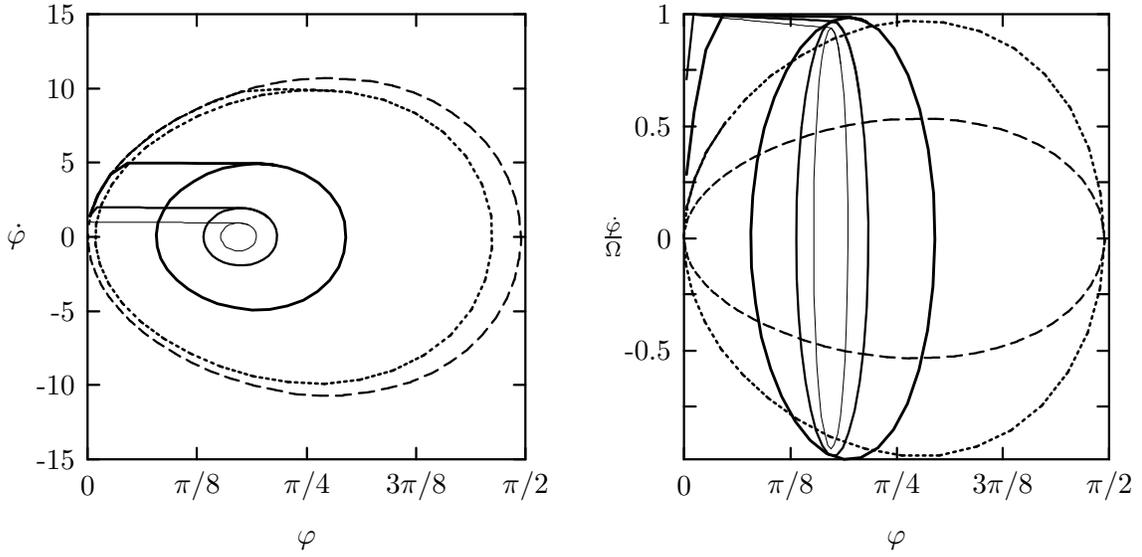
\fi
The particle dynamics can be understood in more detail by looking at the motion
in phase space, $(\varphi,\dot{\varphi})$, see also ref.~\cite{schinner97}.
This is shown in Fig.~\ref{fig: one_part4} for different rotation speeds,
$\Omega$, of the drum, ranging from  1 to 20 Hz and a particle starting
initially at $(\varphi,\dot{\varphi})=(0,0)$. To the left, Fig.~\ref{fig:
one_part4}a, the motion in the variable $\dot{\varphi}$ is shown whereas to the
right, Fig.~\ref{fig: one_part4}b, $\dot{\varphi}$ is made dimensionless by
dividing by $\Omega$. For small values of $\Omega$, the particle initially
sticks to the surface and rotates with the constant value
$\dot{\varphi}=\Omega$, leading to a horizontal line in phase space. For higher
values of $\Omega$, a longer and longer transition period is visible and for
the highest value, $\Omega=20$ Hz, the particle never reaches the value
$\dot{\varphi}=\Omega$ but stays in full-sliding motion all the time. After the
transition period, the particle oscillates around the value
$\varphi=\textrm{arctan}(\mu_w)$ which shows up as circular motion around the
point $(\varphi,\dot{\varphi})=(\textrm{arctan}(\mu_w),0)$ in the phase space
plots. For our initial conditions, the maximum extend of the circular regime
will be $\varphi \in [0^\circ,90^\circ]$ and $\dot{\varphi} \in
[-\Omega,\Omega]$ which is roughly given by the line corresponding to
$\Omega=11$ Hz in Fig.~\ref{fig: one_part4}. Please note that the centrifugal
motion will correspond to a horizontal line at $\dot{\varphi}\equiv\Omega$.

As discussed above, a simple regularized Coulomb friction law, see
Eq.~(\ref{eq: fs}), led to two different kinds of trajectories depending on the
initial condition of the system. For initial amplitudes below $\Delta\varphi$,
the system will oscillate with this amplitude whereas for initial amplitudes
above $\Delta\varphi$, the system dissipates energy to reach a final
oscillation with an amplitude lower than or equal to $\Delta\varphi$. By
considering different friction laws, e.g.\ choosing a different friction
coefficient for the low and high velocity regime, up to four different kinds of
trajectories were found analytically~\cite{schinner97}. The authors could also
partly account for the stick-slip motion and achieved a closer agreement
between theory and experiment.

When a round particle, e.g.\ a sphere, is put in a rotating drum, it will start
to rotate with the same sense of rotation as the drum and one might expect to be
able to measure indirectly the coefficient of rolling friction by a similar
method as described above. Unfortunately, the surface properties in our
numerical model are too smooth and the angle of the particle, measured with
respect to the vertical, will always give a value around zero, regardless of the
angular velocity of the drum or the friction coefficient.

%%%%%%%%%%%%%%%%%%%%%%%%%%%%%%%%%%%%%%%%%%%%%%%%%%%%%%%%%%%%%%%%%%%%%%%%%%%%%%%
\section{Few Rotating Particles in a Two-Dimensional Drum}
\label{sec: few_part}
If more than one round particle is placed in a two-dimensional rotating drum, 
the particle rotations can be frustrated, i.e.\ suppressed, whenever particles
have more than one contact point. The collective behaviour of all particles
will then be very similar to that of a non-rotating particle as described in
section~\ref{sec: one_part} and the agreement becomes better with increasing
particle number.
\ifx\draft\undefined
\else
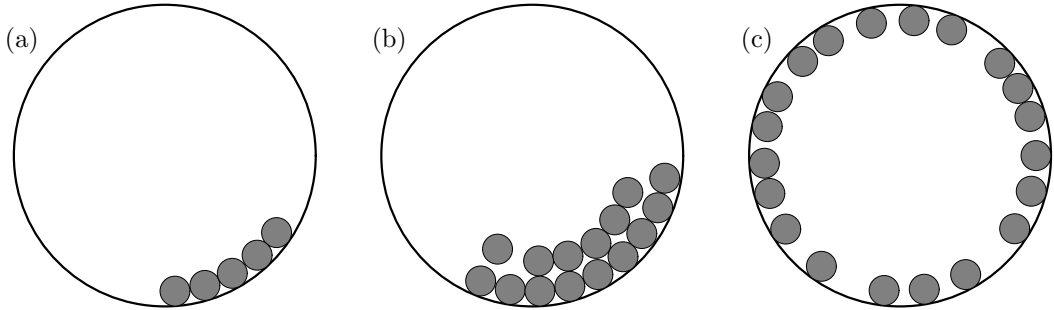
\begin{figure}[t]
  \hbox{%LaTeX picture with EEPIC extensions
\setlength{\unitlength}{0.155pt}
\begin{picture}(850,900)(0,0)
\small
\put(100,700){\makebox(0,0){(a)}}
\thicklines \put(449,418){\circle{738}}
\thinlines
\put(722,230){\shade\circle{ 73.8}}
\put(474, 87){\shade\circle{ 73.8}}
\put(547,101){\shade\circle{ 73.8}}
\put(674,175){\shade\circle{ 73.8}}
\put(614,131){\shade\circle{ 73.8}}
\end{picture} %LaTeX picture with EEPIC extensions
\setlength{\unitlength}{0.155pt}
\begin{picture}(850,900)(0,0)
\small
\put(100,700){\makebox(0,0){(b)}}
\thicklines \put(449,418){\circle{738}}
\thinlines
\put(364,190){\shade\circle{ 73.8}}
\put(464,161){\shade\circle{ 73.8}}
\put(682,328){\shade\circle{ 73.8}}
\put(650,261){\shade\circle{ 73.8}}
\put(395, 91){\shade\circle{ 73.8}}
\put(536,173){\shade\circle{ 73.8}}
\put(772,363){\shade\circle{ 73.8}}
\put(323,112){\shade\circle{ 73.8}}
\put(609,128){\shade\circle{ 73.8}}
\put(670,171){\shade\circle{ 73.8}}
\put(467, 87){\shade\circle{ 73.8}}
\put(540, 99){\shade\circle{ 73.8}}
\put(755,291){\shade\circle{ 73.8}}
\put(604,204){\shade\circle{ 73.8}}
\put(716,228){\shade\circle{ 73.8}}
\end{picture} %LaTeX picture with EEPIC extensions
\setlength{\unitlength}{0.155pt}
\begin{picture}(850,900)(0,0)
\small
\put(100,700){\makebox(0,0){(c)}}
\thicklines \put(449,418){\circle{738}}
\thinlines
\put(769,331){\shade\circle{ 73.8}}
\put(274,699){\shade\circle{ 73.8}}
\put(737,582){\shade\circle{ 73.8}}
\put(729,240){\shade\circle{ 73.8}}
\put(257,148){\shade\circle{ 73.8}}
\put(170,239){\shade\circle{ 73.8}}
\put(781,418){\shade\circle{ 73.8}}
\put(379,742){\shade\circle{ 73.8}}
\put(766,514){\shade\circle{ 73.8}}
\put(211,648){\shade\circle{ 73.8}}
\put(118,400){\shade\circle{ 73.8}}
\put(508, 92){\shade\circle{ 73.8}}
\put(609,128){\shade\circle{ 73.8}}
\put(575,725){\shade\circle{ 73.8}}
\put(410, 89){\shade\circle{ 73.8}}
\put(131,326){\shade\circle{ 73.8}}
\put(483,748){\shade\circle{ 73.8}}
\put(692,644){\shade\circle{ 73.8}}
\put(150,562){\shade\circle{ 73.8}}
\put(125,489){\shade\circle{ 73.8}}
\end{picture} }
  \caption[]{\parbox[t]{0.85\textwidth}{Different motion of few rotating
             particles $N$ for $\Omega=50$ Hz, $\mu=0.75$ and $\mu_w=0.35$: (a) 
             $N=5$, (b) $N=15$ and (c) $N=20$.}}
  \label{fig: few_part1}
\end{figure}
\fi

To investigate the transition to the centrifugal motion in this case, a
two-dimensional MD code is used which includes particle rotations. For the
normal forces, the linear force with a linear damping was used and for the
tangential force, the viscous friction law with a high enough parameter
$\gamma_s$, to ensure that mostly the Coulomb friction law applies, was chosen.
Two different friction coefficients will be used, namely: i) one for
particle-particle collisions denoted by $\mu$ and ii) one for particle-wall
collisions denoted by $\mu_w$ which will be varied independently. All particles
have a diameter of $d=1$ cm and the drum diameter was $D=10$ cm. This gives a
maximum of 28 particles that can form the outer ring in the centrifugal regime
and we limit our investigations to this value in order for all particles to
feel the wall friction.

For only a few number of particles, they will all form a line, shown in
Fig.~\ref{fig: few_part1}a for $N=5$, and the motion will essentially be
one-dimensional. For more particles, where the exact number depends linearly on
the drum to particle diameter ratio, $D/d$, more than one layer or ring of
particles will form, shown in Fig.~\ref{fig: few_part1}b for $N=15$, and the
motion is no longer one-dimensional but two-dimensional. Above a critical
particle number, $N_c$, which depends on $D/d$ as well but also on the friction
coefficient, $\mu$, all particles will form a single ring again but make full
overturns~\cite{scherer}. Such a situation is depicted in Fig.~\ref{fig:
few_part1}c for $N=20$ and will be called the {\em centrifugal regime} in this
case.

First we look at the angle with respect to the vertical averaged over all 
particles, $\langle\varphi\rangle$, as a function of time. For a small number of
particles, it shows a similar behaviour as the angle of a non-rotating particle
which is shown in Fig.~\ref{fig: few_part2}a for the configuration from
Fig.~\ref{fig: few_part1}a. Initially, the five particles form two layers but
reorganize into one layer at $t\approx 1.25$ s which can be seen in
Fig.~\ref{fig: few_part2}a by the change in amplitude. After this
reorganization, oscillations around arctan($\mu_w$), shown as dotted line, will
occur as was the case for one non-rotating particle, see Fig.~\ref{fig:
one_part1}b.
\ifx\draft\undefined
\else
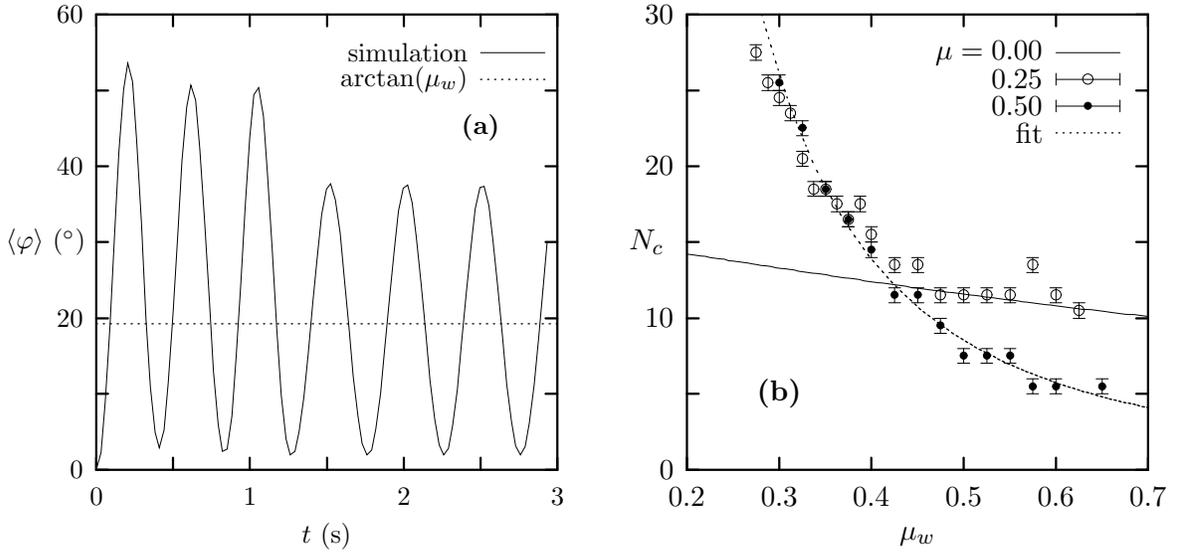
\begin{figure}[t]
  \hbox{\hspace{-0.7cm}% GNUPLOT: LaTeX picture using EEPIC macros
\setlength{\unitlength}{0.225pt}
\begin{picture}(1049,900)(0,50)
\small
\thicklines \path(177,179)(197,179)
\thicklines \path(952,179)(932,179)
\put(155,179){\makebox(0,0)[r]{0}}
\thicklines \path(177,307)(197,307)
\thicklines \path(952,307)(932,307)
\put(155,307){\makebox(0,0)[r]{}}
\thicklines \path(177,434)(197,434)
\thicklines \path(952,434)(932,434)
\put(155,434){\makebox(0,0)[r]{20}}
\thicklines \path(177,562)(197,562)
\thicklines \path(952,562)(932,562)
\put(155,562){\makebox(0,0)[r]{}}
\thicklines \path(177,690)(197,690)
\thicklines \path(952,690)(932,690)
\put(155,690){\makebox(0,0)[r]{40}}
\thicklines \path(177,817)(197,817)
\thicklines \path(952,817)(932,817)
\put(155,817){\makebox(0,0)[r]{}}
\thicklines \path(177,945)(197,945)
\thicklines \path(952,945)(932,945)
\put(155,945){\makebox(0,0)[r]{60}}
\thicklines \path(177,179)(177,199)
\thicklines \path(177,945)(177,925)
\put(177,134){\makebox(0,0){0}}
\thicklines \path(306,179)(306,199)
\thicklines \path(306,945)(306,925)
\put(306,134){\makebox(0,0){}}
\thicklines \path(435,179)(435,199)
\thicklines \path(435,945)(435,925)
\put(435,134){\makebox(0,0){1}}
\thicklines \path(565,179)(565,199)
\thicklines \path(565,945)(565,925)
\put(565,134){\makebox(0,0){}}
\thicklines \path(694,179)(694,199)
\thicklines \path(694,945)(694,925)
\put(694,134){\makebox(0,0){2}}
\thicklines \path(823,179)(823,199)
\thicklines \path(823,945)(823,925)
\put(823,134){\makebox(0,0){}}
\thicklines \path(952,179)(952,199)
\thicklines \path(952,945)(952,925)
\put(952,134){\makebox(0,0){3}}
\thicklines \path(177,179)(952,179)(952,945)(177,945)(177,179)
\put(25,562){\makebox(0,0)[l]{\shortstack{$\langle\varphi\rangle$ ($^\circ$)}}}
\put(564,67){\makebox(0,0){$t$ (s)}}
\put(823,754){\makebox(0,0){\bf (a)}}
\put(801,881){\makebox(0,0)[r]{simulation}}
\thinlines \path(823,881)(931,881)
\thinlines \path(177,179)(177,179)(185,208)(192,292)(200,420)(207,569)(215,713)(223,819)(230,863)(238,832)(245,732)(253,594)(260,434)(268,318)(276,242)(283,216)(291,246)(298,325)(306,441)(314,574)(321,698)(329,789)(336,826)(344,801)(351,720)(359,600)(367,467)(374,345)(382,255)(389,210)(397,214)(405,269)(412,366)(420,492)(427,624)(435,738)(442,810)(450,822)(458,774)(465,674)(473,546)(480,415)(488,304)(496,231)(503,204)(511,210)(518,243)(526,301)(533,376)(541,460)(549,542)
\thinlines \path(549,542)(556,609)(564,651)(571,660)(579,633)(587,577)(594,501)(602,417)(609,336)(617,269)(624,223)(632,204)(640,212)(647,248)(655,308)(662,384)(670,468)(678,549)(685,614)(693,653)(700,658)(708,629)(715,571)(723,493)(731,409)(738,330)(746,264)(753,221)(761,204)(769,215)(776,253)(784,314)(791,392)(799,476)(807,555)(814,618)(822,654)(829,656)(837,624)(844,564)(852,486)(860,402)(867,323)(875,260)(882,219)(890,204)(898,218)(905,258)(913,321)(920,399)(928,483)
\thinlines \path(928,483)(935,562)
\put(801,836){\makebox(0,0)[r]{arctan($\mu_w$)}}
\dottedline{12}(823,836)(931,836)
\dottedline{12}(177,425)(952,425)
\end{picture} \hfill \hspace{-0.7cm}% GNUPLOT: LaTeX picture using EEPIC macros
%\setlength{\unitlength}{0.240900pt}
%\begin{picture}(1500,900)(0,0)
\setlength{\unitlength}{0.225pt}
\begin{picture}(1049,900)(0,50)
%\tenrm
\thicklines \path(177,179)(197,179)
\thicklines \path(952,179)(932,179)
\put(155,179){\makebox(0,0)[r]{0}}
\thicklines \path(177,307)(197,307)
\thicklines \path(952,307)(932,307)
\put(155,307){\makebox(0,0)[r]{}}
\thicklines \path(177,434)(197,434)
\thicklines \path(952,434)(932,434)
\put(155,434){\makebox(0,0)[r]{10}}
\thicklines \path(177,562)(197,562)
\thicklines \path(952,562)(932,562)
\put(155,562){\makebox(0,0)[r]{}}
\thicklines \path(177,690)(197,690)
\thicklines \path(952,690)(932,690)
\put(155,690){\makebox(0,0)[r]{20}}
\thicklines \path(177,817)(197,817)
\thicklines \path(952,817)(932,817)
\put(155,817){\makebox(0,0)[r]{}}
\thicklines \path(177,945)(197,945)
\thicklines \path(952,945)(932,945)
\put(155,945){\makebox(0,0)[r]{30}}
\thicklines \path(177,179)(177,199)
\thicklines \path(177,945)(177,925)
\put(177,134){\makebox(0,0){0.2}}
\thicklines \path(332,179)(332,199)
\thicklines \path(332,945)(332,925)
\put(332,134){\makebox(0,0){0.3}}
\thicklines \path(487,179)(487,199)
\thicklines \path(487,945)(487,925)
\put(487,134){\makebox(0,0){0.4}}
\thicklines \path(642,179)(642,199)
\thicklines \path(642,945)(642,925)
\put(642,134){\makebox(0,0){0.5}}
\thicklines \path(797,179)(797,199)
\thicklines \path(797,945)(797,925)
\put(797,134){\makebox(0,0){0.6}}
\thicklines \path(952,179)(952,199)
\thicklines \path(952,945)(952,925)
\put(952,134){\makebox(0,0){0.7}}
\thicklines \path(177,179)(952,179)(952,945)(177,945)(177,179)
\put(80,562){\makebox(0,0)[l]{\shortstack{$N_c$}}}
\put(564,67){\makebox(0,0){$\mu_w$}}
\put(332,307){\makebox(0,0){\bf (b)}}
\put(775,881){\makebox(0,0)[r]{$\mu=0.00$}}
\thinlines \path(797,881)(905,881)
\thinlines \path(177,542)(177,542)(185,541)(193,540)(200,539)(208,538)(216,536)(224,535)(232,534)(240,533)(247,531)(255,530)(263,529)(271,528)(279,527)(287,525)(294,524)(302,523)(310,522)(318,521)(326,519)(334,518)(341,517)(349,516)(357,515)(365,513)(373,512)(381,511)(388,510)(396,509)(404,508)(412,507)(420,505)(428,504)(435,503)(443,502)(451,501)(459,500)(467,499)(474,497)(482,496)(490,495)(498,494)(506,493)(514,492)(521,491)(529,490)(537,489)(545,488)(553,486)(561,485)
\thinlines \path(561,485)(568,484)(576,483)(584,482)(592,481)(600,480)(608,479)(615,478)(623,477)(631,476)(639,475)(647,474)(655,473)(662,472)(670,471)(678,470)(686,469)(694,468)(701,467)(709,466)(717,465)(725,464)(733,463)(741,462)(748,461)(756,460)(764,459)(772,458)(780,457)(788,456)(795,455)(803,454)(811,453)(819,452)(827,451)(835,450)(842,449)(850,449)(858,448)(866,447)(874,446)(882,445)(889,444)(897,443)(905,442)(913,441)(921,440)(929,440)(936,439)(944,438)(952,437)
\put(775,836){\makebox(0,0)[r]{$0.25$}}
\thinlines \path(797,836)(905,836)
\thinlines \path(797,846)(797,826)
\thinlines \path(905,846)(905,826)
\thinlines \path(293,868)(293,894)
\thinlines \path(283,868)(303,868)
\thinlines \path(283,894)(303,894)
\thinlines \path(313,817)(313,843)
\thinlines \path(303,817)(323,817)
\thinlines \path(303,843)(323,843)
\thinlines \path(332,792)(332,817)
\thinlines \path(322,792)(342,792)
\thinlines \path(322,817)(342,817)
\thinlines \path(351,766)(351,792)
\thinlines \path(341,766)(361,766)
\thinlines \path(341,792)(361,792)
\thinlines \path(371,690)(371,715)
\thinlines \path(361,690)(381,690)
\thinlines \path(361,715)(381,715)
\thinlines \path(390,639)(390,664)
\thinlines \path(380,639)(400,639)
\thinlines \path(380,664)(400,664)
\thinlines \path(410,639)(410,664)
\thinlines \path(400,639)(420,639)
\thinlines \path(400,664)(420,664)
\thinlines \path(429,613)(429,639)
\thinlines \path(419,613)(439,613)
\thinlines \path(419,639)(439,639)
\thinlines \path(448,588)(448,613)
\thinlines \path(438,588)(458,588)
\thinlines \path(438,613)(458,613)
\thinlines \path(468,613)(468,639)
\thinlines \path(458,613)(478,613)
\thinlines \path(458,639)(478,639)
\thinlines \path(487,562)(487,588)
\thinlines \path(477,562)(497,562)
\thinlines \path(477,588)(497,588)
\thinlines \path(526,511)(526,536)
\thinlines \path(516,511)(536,511)
\thinlines \path(516,536)(536,536)
\thinlines \path(565,511)(565,536)
\thinlines \path(555,511)(575,511)
\thinlines \path(555,536)(575,536)
\thinlines \path(603,460)(603,485)
\thinlines \path(593,460)(613,460)
\thinlines \path(593,485)(613,485)
\thinlines \path(642,460)(642,485)
\thinlines \path(632,460)(652,460)
\thinlines \path(632,485)(652,485)
\thinlines \path(681,460)(681,485)
\thinlines \path(671,460)(691,460)
\thinlines \path(671,485)(691,485)
\thinlines \path(720,460)(720,485)
\thinlines \path(710,460)(730,460)
\thinlines \path(710,485)(730,485)
\thinlines \path(758,511)(758,536)
\thinlines \path(748,511)(768,511)
\thinlines \path(748,536)(768,536)
\thinlines \path(797,460)(797,485)
\thinlines \path(787,460)(807,460)
\thinlines \path(787,485)(807,485)
\thinlines \path(836,434)(836,460)
\thinlines \path(826,434)(846,434)
\thinlines \path(826,460)(846,460)
\put(293,881){\circle{18}}
\put(313,830){\circle{18}}
\put(332,805){\circle{18}}
\put(351,779){\circle{18}}
\put(371,702){\circle{18}}
\put(390,651){\circle{18}}
\put(410,651){\circle{18}}
\put(429,626){\circle{18}}
\put(448,600){\circle{18}}
\put(468,626){\circle{18}}
\put(487,575){\circle{18}}
\put(526,524){\circle{18}}
\put(565,524){\circle{18}}
\put(603,473){\circle{18}}
\put(642,473){\circle{18}}
\put(681,473){\circle{18}}
\put(720,473){\circle{18}}
\put(758,524){\circle{18}}
\put(797,473){\circle{18}}
\put(836,447){\circle{18}}
\put(851,836){\circle{18}}
\put(775,791){\makebox(0,0)[r]{$0.50$}}
\thinlines \path(797,791)(905,791)
\thinlines \path(797,801)(797,781)
\thinlines \path(905,801)(905,781)
\thinlines \path(332,817)(332,843)
\thinlines \path(322,817)(342,817)
\thinlines \path(322,843)(342,843)
\thinlines \path(371,741)(371,766)
\thinlines \path(361,741)(381,741)
\thinlines \path(361,766)(381,766)
\thinlines \path(410,639)(410,664)
\thinlines \path(400,639)(420,639)
\thinlines \path(400,664)(420,664)
\thinlines \path(448,588)(448,613)
\thinlines \path(438,588)(458,588)
\thinlines \path(438,613)(458,613)
\thinlines \path(487,536)(487,562)
\thinlines \path(477,536)(497,536)
\thinlines \path(477,562)(497,562)
\thinlines \path(526,460)(526,485)
\thinlines \path(516,460)(536,460)
\thinlines \path(516,485)(536,485)
\thinlines \path(565,460)(565,485)
\thinlines \path(555,460)(575,460)
\thinlines \path(555,485)(575,485)
\thinlines \path(603,409)(603,434)
\thinlines \path(593,409)(613,409)
\thinlines \path(593,434)(613,434)
\thinlines \path(642,358)(642,383)
\thinlines \path(632,358)(652,358)
\thinlines \path(632,383)(652,383)
\thinlines \path(681,358)(681,383)
\thinlines \path(671,358)(691,358)
\thinlines \path(671,383)(691,383)
\thinlines \path(720,358)(720,383)
\thinlines \path(710,358)(730,358)
\thinlines \path(710,383)(730,383)
\thinlines \path(758,307)(758,332)
\thinlines \path(748,307)(768,307)
\thinlines \path(748,332)(768,332)
\thinlines \path(797,307)(797,332)
\thinlines \path(787,307)(807,307)
\thinlines \path(787,332)(807,332)
\thinlines \path(875,307)(875,332)
\thinlines \path(865,307)(885,307)
\thinlines \path(865,332)(885,332)
\put(332,830){\circle*{12}}
\put(371,754){\circle*{12}}
\put(410,651){\circle*{12}}
\put(448,600){\circle*{12}}
\put(487,549){\circle*{12}}
\put(526,473){\circle*{12}}
\put(565,473){\circle*{12}}
\put(603,422){\circle*{12}}
\put(642,371){\circle*{12}}
\put(681,371){\circle*{12}}
\put(720,371){\circle*{12}}
\put(758,319){\circle*{12}}
\put(797,319){\circle*{12}}
\put(875,319){\circle*{12}}
\put(851,791){\circle*{12}}
\put(775,746){\makebox(0,0)[r]{fit}}
\thinlines \dottedline{12}(797,746)(905,746)
\thinlines
\dottedline{12}(303,945)(310,919)(318,892)(326,865)(334,840)(341,817)(349,794)(357,773)(365,753)(373,734)(381,715)(388,698)(396,681)(404,665)(412,650)(420,636)(428,622)(435,609)(443,596)(451,584)(459,572)(467,561)(474,551)(482,540)(490,530)(498,521)(506,512)(514,503)(521,495)(529,487)(537,479)(545,471)(553,464)(561,457)(568,450)(576,444)(584,438)(592,432)(600,426)(608,420)(615,415)(623,409)(631,404)(639,399)(647,394)(655,390)(662,385)(670,381)(678,377)(686,373)(694,369)
\dottedline{12}(694,369)(701,365)(709,361)(717,357)(725,354)(733,350)(741,347)(748,344)(756,341)(764,338)(772,335)(780,332)(788,329)(795,326)(803,323)(811,321)(819,318)(827,316)(835,313)(842,311)(850,309)(858,307)(866,304)(874,302)(882,300)(889,298)(897,296)(905,294)(913,292)(921,291)(929,289)(936,287)(944,285)(952,284)
\end{picture} }
  \caption[]{\parbox[t]{0.85\textwidth}{(a) Average angle of 5 rotating
             particles for $\Omega=50$ Hz, $\mu=0.75$ and $\mu_w=0.35$ and (b)
             critical particle number for the transition to the centrifugal
             regime as function of wall friction for three different
             particle-particle friction coefficients $\mu$. The theoretical
	     curve corresponds to the case $\mu=0.00$.}}
  \label{fig: few_part2}
\end{figure}
\fi

To investigate the parameter regimes of the different motions depicted in
Fig.~\ref{fig: few_part1}, especially the transition to the centrifugal regime,
we determine the minimal number of particles, $N_c$, which are necessary to
reach it. For an inter-particle friction coefficient of $\mu=0.0$, all
particles will form a single ring since piles cannot be supported. The particle
configuration will fluctuate around a mean angle of
$\langle\varphi\rangle=\textrm{arctan}(\mu_w)$, see Fig.~\ref{fig: few_part2}a.
As discussed in the preceding section, a necessary condition for overturns is
that the center of mass of the top-most particle reaches an angle of more than
$\frac{\pi}{2}$. The angle $\varphi$ is measured from the vertical pointing 
downwards again, as shown in Fig.~\ref{fig: one_part1}a. Assuming that
all particles are forming a single block and are touching their neighbours, the
center of mass of the top-most particle of a configuration of $N$ particles is
at an angle of $\textrm{arctan}(\mu_w)+\frac{d}{D}(N-\frac{1}{2})$. This
gives as necessary condition for overturns
\[ \textrm{arctan}(\mu_w) + \frac{d}{D}(N-\frac{1}{2})\, >\, \frac{\pi}{2} \]
which when solved for the critical particle number, $N_c$, gives
\beq
  N_c > \frac{D}{d} (\frac{\pi}{2}-\textrm{arctan}(\mu_w))\, +\, \frac{1}{2} \ .
  \label{eq: n_c}
\eeq

This analytic expression for $N_c$, derived for an inter-particle friction
coefficient of $\mu=0.0$, is shown in Fig.~\ref{fig: few_part2}b as function of
the wall friction coefficient, $\mu_w$, as solid line. Also shown are the
numerical results for $\mu = 0.25$ ($\circ$) and $\mu = 0.50$ ($\bullet$) with
a dotted line as least-square power law fit to the data points for the latter
case to guide the eye. The differences with the analytic curve can be explained
in the following way: If the inter-particle friction is sufficiently high,
already two rotating particles can behave as a non-rotating particle in the
two-dimensional drum. From the preceding section, one recalls that overturns
are possible for one non-rotating particle if $\mu_w \gtrsim 0.65$. Therefore,
the critical particle number will {\em decrease} with increasing inter-particle
friction for a high wall friction coefficient and we get $\lim_{\mu \rightarrow
1}N_c=2$ for $\mu_w \gtrsim 0.65$. This is in perfect agreement with the numerical
data where the data points for $\mu = 0.25$ lie very close to the analytic
curve for $\mu_w > 0.45$ and $N_c$ decreases with increasing $\mu$ giving a
value of $N_c \approx 5$ for $\mu=0.5$ and $\mu_w=0.65$.

For low wall friction coefficients, a larger number of particles is needed to
reach the centrifugal regime since the value of $\textrm{arctan}(\mu_w)$ is
small. If $\mu>0$, the particles can be arranged in more than one layer which
gives a lower height for the top-most particle as compared to the one layer
case discussed above. Consequently, $N_c$ will {\em increase} with increasing
inter-particle friction for a low wall friction coefficient since the layer
structure becomes more stable which is supported by the numerical data.

%%%%%%%%%%%%%%%%%%%%%%%%%%%%%%%%%%%%%%%%%%%%%%%%%%%%%%%%%%%%%%%%%%%%%%%%%%%%%%%
\section{Conclusions}
We investigated analytically the motion of a non-rotating particle in a
rotating drum in the full-sliding limit. The angle of the particle's center of
mass oscillates around a mean value, given by the particle-wall Coulomb
friction  coefficient. The one-dimensional description breaks down for angles
above 90$^\circ$, corresponding to free flights of the particle. In order to
study the full dynamics, including the transition from sticking to sliding
motion for low velocities of the contact surfaces, we simulated the system
using molecular dynamics. Full agreement with the analytic results in the
full-sliding case was achieved and the general trend of the frequency
dependence of the average particle angle observed in experiments could also be
reproduced. The phase diagram of the transition to the centrifugal motion
revealed three different regimes.

We also found that the collective motion of a few rotating spheres in a
two-dimensional rotating drum is, to a certain extend, similar to the motion of
one non-rotating particle in a drum due to the frustrated particle rotations.
We derived an analytic expression for the minimal number of particles to obtain
centrifugal motion in the limit of negligible inter-particle friction. It
compares well with the experimental findings and the change when inter-particle
friction is added can be understood as well.

\section*{Acknowledgments}
We gratefully acknowledge fruitful and helpful discussions with C.\ Dury, A.\ 
Betat, K.\ Kassner, I.\ Rehberg and A.\ Schinner. Special thanks go to M.\ 
Scherer for sharing his experimental ideas with us and to the {\em Deutsche
Forschungsgemeinschaft} for financial support.

%%%%%%%% REFERENCES %%%%%%%%

%\bibliographystyle{../prsty}
%\bibliography{habil}

\begin{thebibliography}{10}

\bibitem{jaeger96}
H.~M. Jaeger, S.~R. Nagel, and R.~P. Behringer, Physics Today {\bf 4},  32
  (1996).

\bibitem{rajchenbach90}
J. Rajchenbach, Phys. Rev. Lett. {\bf 65},  2221  (1990).

\bibitem{dury97d}
C.~M. Dury, G.~H. Ristow, J.~L. Moss, and M. Nakagawa, Phys. Rev. E {\bf 57},
  4491  (1998).

\bibitem{nityanand86}
N. Nityanand, B. Manley, and H. Henein, Metall. Trans. B {\bf 17},  247
  (1986).

\bibitem{caponeri95}
M. Caponeri, S. Douady, S. Fauve, and C. Laroche,  in {\em Mobile Particulate
  Systems}, edited by E. Guazzelli and L. Oger (Kluwer, Dordrecht, 1995), p.\
  331.

\bibitem{bhushan95}
B. Bhushan, J.~N. Israelachvill, and U. Landman, Nature {\bf 374},  607
  (1995).

\bibitem{heslot94}
F. Heslot, T. Baumberger, B. Perrin, B. Caroli, and C. Caroli, Phys. Rev. E
  {\bf 49},  4973  (1994).

\bibitem{elmer97}
F.-J. Elmer, J. Phys. A: Math. Gen. {\bf 30},  6057  (1997).

\bibitem{betat97}
A. Betat and I. Rehberg,  in {\em Friction, Arching, Contact Dynamics}, edited
  by D.~E. Wolf and P. Grassberger (World Scientific, Singapore, 1997), p.\
  301.

\bibitem{kassner}
K. Kassner, private communications, 1996.

\bibitem{ristow94b}
G.~H. Ristow,  in {\em Annual Reviews of Computational Physics I}, edited by D.
  Stauffer (World Scientific, Singapore, 1994), p.\ 275.

\bibitem{schinner97}
A. Schinner and K. Kassner,  in {\em Friction, Arching, Contact Dynamics},
  edited by D.~E. Wolf and P. Grassberger (World Scientific, Singapore, 1997),
  p.\ 305.

\bibitem{scherer}
M. Scherer, private communications, 1996.

\end{thebibliography}

%%%%%%%%%%%%%%%%%%%%%%%%%%%%

\ifx\draft\undefined
%%%%%%%%%%%%%%%%%%%%%%%%%%%%%%%%%%%%%%%%%%%%%%%%%%%%%%%%%%%%%%%%%%%%%%%%%%%%
%  FIGURES
%%%%%%%%%%%%%%%%%%%%%%%%%%%%%%%%%%%%%%%%%%%%%%%%%%%%%%%%%%%%%%%%%%%%%%%%%%%%
\newpage
\subsection*{Figure Captions}
\begin{description}
\item[Figure~\ref{fig: one_part1}:] (a) Sketch of the quasi one-dimensional
             setup and (b) evolution of the angle as function of time in the
             simulation for $\mu_w=0.47$.
\item[Figure~\ref{fig: one_part2}:] (a) Comparison of the analytic
             expression for the maximum amplitude with the simulation and (b) 
             frequency dependence of the average angle for the experiment 
             ($\bullet$) and the simulation using $\mu_w=0.47$ ($\circ$).
\item[Figure~\ref{fig: one_part3}:] (a) Transition frequency to the
             centrifugal regime as function of friction coefficient and (b)
             angle of the particle as function of time for $\mu_w=0.665$ and
             $\Omega=43$ Hz (regime II).
\item[Figure~\ref{fig: one_part4}:] Phase space plot for different angular 
             frequencies $\Omega\!=\!1($---$),2(${\bf ---}$),5 $({\bf ---})$,
	     11(\cdot \cdot \cdot),20($-- --) Hz, (a) unnormalized and (b) 
	     normalized by $\Omega$. 
\item[Figure~\ref{fig: few_part1}:] different motion of few rotating
             particles $N$ for $\Omega=50$ Hz, $\mu=0.75$ and $\mu_w=0.35$: (a) 
             $N=5$, (b) $N=15$ and (c) $N=20$
\item[Figure~\ref{fig: few_part2}:] (a) Average angle of 5 rotating
             particles for $\Omega=50$ Hz, $\mu=0.75$ and $\mu_w=0.35$ and (b)
             critical particle number for the transition to the centrifugal
             regime as function of wall friction for three different
             particle-particle friction coefficients $\mu$. The theoretical
	     curve corresponds to the case $\mu=0.00$.
\end{description}

\newpage

\begin{figure}[htb]
  \hbox{\hspace{-0.7cm}\setlength{\unitlength}{0.0006in}
%{\renewcommand{\dashlinestretch}{30}
\begin{picture}(5500,3200)(-300,-850)

\Thicklines

\put(1927,1927){\circle{3794}}
\put(3062,912){\circle{750}}
\put(3062,912){\shade\circle{750}}

\blacken\path(4207.565,2367.785)(4029.000,2539.000)(4105.986,2303.897)(4118.443,2396.789)(4207.565,2367.785)
\blacken\path(669.000,1737.000)(729.000,1497.000)(789.000,1737.000)(729.000,1665.000)(669.000,1737.000)

\path(1927,1927)(1927,52)
\path(1927,1927)(3356,652)
\path(729,2389)(729,1497)

\path(2327,2327)(3462,1312)
\path(2276,2270)(2378,2384)
\path(3411,1255)(3513,1369)
\put(2874,1857){\makebox(0,0)[lb]{$R$}}

\put(4374,1857){\makebox(0,0)[lb]{$\Omega$}}
\put(3034.909,1913.773){\arc{2348.724}{5.7218}{6.7946}}

\put(2080,1335){\makebox(0,0)[lb]{$\varphi$}}
\put(2114.500,1428.250){\arc{676.041}{6.2277}{8.4420}}

\put(950,1857){\makebox(0,0)[lb]{$\vec{g}$}}

\put(3900,3400){\makebox(0,0)[lb]{\bf (a)}}

\end{picture} \hfill \hspace{-0.7cm}% GNUPLOT: LaTeX picture using EEPIC macros
\setlength{\unitlength}{0.225pt}
\begin{picture}(1049,900)(0,50)
\small
\thicklines \path(177,179)(197,179)
\thicklines \path(952,179)(932,179)
\put(155,179){\makebox(0,0)[r]{0}}
\thicklines \path(177,307)(197,307)
\thicklines \path(952,307)(932,307)
\put(155,307){\makebox(0,0)[r]{}}
\thicklines \path(177,434)(197,434)
\thicklines \path(952,434)(932,434)
\put(155,434){\makebox(0,0)[r]{10}}
\thicklines \path(177,562)(197,562)
\thicklines \path(952,562)(932,562)
\put(155,562){\makebox(0,0)[r]{}}
\thicklines \path(177,690)(197,690)
\thicklines \path(952,690)(932,690)
\put(155,690){\makebox(0,0)[r]{20}}
\thicklines \path(177,817)(197,817)
\thicklines \path(952,817)(932,817)
\put(155,817){\makebox(0,0)[r]{}}
\thicklines \path(177,945)(197,945)
\thicklines \path(952,945)(932,945)
\put(155,945){\makebox(0,0)[r]{30}}
\thicklines \path(177,179)(177,199)
\thicklines \path(177,945)(177,925)
\put(177,134){\makebox(0,0){0}}
\thicklines \path(306,179)(306,199)
\thicklines \path(306,945)(306,925)
\put(306,134){\makebox(0,0){}}
\thicklines \path(435,179)(435,199)
\thicklines \path(435,945)(435,925)
\put(435,134){\makebox(0,0){1}}
\thicklines \path(565,179)(565,199)
\thicklines \path(565,945)(565,925)
\put(565,134){\makebox(0,0){}}
\thicklines \path(694,179)(694,199)
\thicklines \path(694,945)(694,925)
\put(694,134){\makebox(0,0){2}}
\thicklines \path(823,179)(823,199)
\thicklines \path(823,945)(823,925)
\put(823,134){\makebox(0,0){}}
\thicklines \path(952,179)(952,199)
\thicklines \path(952,945)(952,925)
\put(952,134){\makebox(0,0){3}}
\thicklines \path(177,179)(952,179)(952,945)(177,945)(177,179)
\put(45,562){\makebox(0,0)[l]{\shortstack{$\varphi$ ($^\circ$)}}}
\put(564,50){\makebox(0,0){$t$ (s)}}
\put(875,562){\makebox(0,0){\bf (b)}}
\put(775,371){\makebox(0,0)[r]{simulation}}
\thinlines \path(797,371)(905,371)
\thinlines \path(179,186)(179,186)(182,202)(184,219)(187,236)(189,253)(191,270)(194,287)(196,304)(199,321)(201,337)(203,354)(206,371)(208,387)(211,404)(213,420)(215,437)(218,453)(220,469)(223,486)(225,502)(227,518)(230,535)(232,550)(235,567)(237,584)(239,601)(242,617)(244,633)(247,645)(249,662)(251,678)(254,694)(256,711)(259,726)(261,740)(263,756)(266,773)(268,789)(270,802)(273,819)(275,835)(278,849)(280,862)(282,878)(285,888)(287,900)(290,909)(292,914)(294,919)(297,923)
\thinlines \path(297,923)(299,923)(302,919)(304,914)(306,906)(309,897)(311,883)(314,871)(316,857)(318,844)(321,828)(323,811)(326,797)(328,781)(330,769)(333,756)(335,747)(338,739)(340,731)(342,726)(345,726)(347,726)(350,728)(352,735)(354,740)(357,752)(359,761)(362,773)(364,785)(366,802)(369,819)(371,831)(374,849)(376,862)(378,875)(381,888)(383,900)(386,909)(388,914)(390,919)(393,923)(395,923)(398,919)(400,914)(402,906)(405,897)(407,888)(410,875)(412,857)(414,845)(417,828)
\thinlines \path(417,828)(419,814)(421,799)(424,785)(426,773)(429,759)(431,747)(433,740)(436,731)(438,726)(441,726)(443,726)(445,726)(448,731)(450,740)(453,747)(455,761)(457,773)(460,785)(462,802)(465,814)(467,831)(469,845)(472,862)(474,875)(477,888)(479,897)(481,906)(484,914)(486,919)(489,923)(491,923)(493,919)(496,914)(498,909)(501,897)(503,888)(505,875)(508,862)(510,845)(513,831)(515,814)(517,802)(520,785)(522,773)(525,761)(527,749)(529,740)(532,735)(534,726)(537,726)
\thinlines \path(537,726)(539,726)(541,726)(544,731)(546,740)(549,747)(551,759)(553,769)(556,785)(558,799)(560,814)(563,828)(565,845)(568,857)(570,875)(572,887)(575,897)(577,906)(580,914)(582,919)(584,923)(587,923)(589,919)(592,914)(594,909)(596,900)(599,888)(601,875)(604,862)(606,849)(608,831)(611,819)(613,802)(616,789)(618,773)(620,761)(623,752)(625,740)(628,735)(630,728)(632,726)(635,726)(637,726)(640,731)(642,739)(644,747)(647,756)(649,768)(652,781)(654,797)(656,811)
\thinlines \path(656,811)(659,828)(661,840)(664,857)(666,871)(668,883)(671,897)(673,906)(676,914)(678,919)(680,923)(683,923)(685,919)(688,914)(690,909)(692,900)(695,892)(697,880)(700,866)(702,849)(704,837)(707,819)(709,802)(712,790)(714,776)(716,764)(719,752)(721,744)(724,735)(726,731)(728,726)(731,726)(733,726)(735,731)(738,739)(740,747)(743,756)(745,768)(747,781)(750,794)(752,811)(755,823)(757,840)(759,857)(762,871)(764,883)(767,897)(769,906)(771,914)(774,917)(776,923)
\thinlines \path(776,923)(779,923)(781,919)(783,917)(786,909)(788,900)(791,892)(793,880)(795,866)(798,852)(800,837)(803,819)(805,806)(807,790)(810,776)(812,764)(815,752)(817,744)(819,735)(822,731)(824,726)(827,726)(829,726)(831,731)(834,735)(836,744)(839,756)(841,764)(843,781)(846,794)(848,807)(851,823)(853,840)(855,854)(858,866)(860,883)(863,892)(865,906)(867,911)(870,917)(872,923)(875,923)(877,923)(879,917)(882,911)(884,902)(886,892)(889,880)(891,866)(894,854)(896,837)
\put(775,320){\makebox(0,0)[r]{arctan($\mu_w$)}}
\thinlines
\dottedline{12}(797,320)(905,320)
\dottedline{12}(177,822)(952,822)
\end{picture} }
  \caption{}
  \label{fig: one_part1}
\end{figure}

\vspace{3cm}

\begin{figure}[htb]
  \hbox{\hspace{-0.7cm}% GNUPLOT: LaTeX picture using EEPIC macros
\setlength{\unitlength}{0.225pt}
\begin{picture}(1049,900)(0,50)
\small
\thicklines \path(199,179)(219,179)
\thicklines \path(952,179)(932,179)
\put(177,179){\makebox(0,0)[r]{0}}
\thicklines \path(199,300)(219,300)
\thicklines \path(952,300)(932,300)
%\put(177,300){\makebox(0,0)[r]{30}}
\thicklines \path(199,421)(219,421)
\thicklines \path(952,421)(932,421)
\put(177,421){\makebox(0,0)[r]{60}}
\thicklines \path(199,542)(219,542)
\thicklines \path(952,542)(932,542)
%\put(177,542){\makebox(0,0)[r]{90}}
\thicklines \path(199,663)(219,663)
\thicklines \path(952,663)(932,663)
\put(177,663){\makebox(0,0)[r]{120}}
\thicklines \path(199,784)(219,784)
\thicklines \path(952,784)(932,784)
%\put(177,784){\makebox(0,0)[r]{150}}
\thicklines \path(199,905)(219,905)
\thicklines \path(952,905)(932,905)
\put(177,905){\makebox(0,0)[r]{180}}
\thicklines \path(199,179)(199,199)
\thicklines \path(199,945)(199,925)
\put(199,134){\makebox(0,0){0}}
\thicklines \path(299,179)(299,199)
\thicklines \path(299,945)(299,925)
%\put(299,134){\makebox(0,0){0.1}}
\thicklines \path(400,179)(400,199)
\thicklines \path(400,945)(400,925)
\put(400,134){\makebox(0,0){0.2}}
\thicklines \path(500,179)(500,199)
\thicklines \path(500,945)(500,925)
%\put(500,134){\makebox(0,0){0.3}}
\thicklines \path(601,179)(601,199)
\thicklines \path(601,945)(601,925)
\put(601,134){\makebox(0,0){0.4}}
\thicklines \path(701,179)(701,199)
\thicklines \path(701,945)(701,925)
%\put(701,134){\makebox(0,0){0.5}}
\thicklines \path(801,179)(801,199)
\thicklines \path(801,945)(801,925)
\put(801,134){\makebox(0,0){0.6}}
\thicklines \path(902,179)(902,199)
\thicklines \path(902,945)(902,925)
%\put(902,134){\makebox(0,0){0.7}}
\thicklines \path(199,179)(952,179)(952,945)(199,945)(199,179)
\put(25,542){\makebox(0,0)[l]{\shortstack{$\Delta \varphi$ ($^\circ$)}}}
\put(575,50){\makebox(0,0){$\mu_w$}}
\put(877,421){\makebox(0,0){\bf (a)}}
\put(478,824){\makebox(0,0)[r]{analytic}}
\thinlines \path(500,824)(608,824)
\thinlines \path(199,179)(199,179)(223,190)(246,201)(270,212)(294,223)(317,234)(341,245)(365,256)(389,267)(412,279)(436,290)(460,302)(483,314)(507,326)(531,339)(554,351)(578,365)(602,378)(625,393)(649,408)(673,424)(697,440)(720,459)(744,479)(768,501)(791,526)(815,555)(839,591)(862,639)(886,711)(910,933)
\put(478,779){\makebox(0,0)[r]{simulation}}
\put(249,203){\circle{18}}
\put(299,226){\circle{18}}
\put(350,250){\circle{18}}
\put(400,274){\circle{18}}
\put(450,299){\circle{18}}
\put(500,325){\circle{18}}
\put(550,351){\circle{18}}
\put(601,380){\circle{18}}
\put(651,410){\circle{18}}
\put(701,446){\circle{18}}
\put(751,486){\circle{18}}
\put(801,541){\circle{18}}
\put(852,618){\circle{18}}
\put(872,671){\circle{18}}
\put(882,711){\circle{18}}
\put(892,778){\circle{18}}
\put(554,779){\circle{18}}
\thinlines
\dottedline{12}(199,542)(952,542)
\end{picture} \hfill \hspace{-0.7cm}% GNUPLOT: LaTeX picture using EEPIC macros
\setlength{\unitlength}{0.225pt}
\begin{picture}(1049,900)(0,50)
\small
\thicklines \path(177,179)(197,179)
\thicklines \path(952,179)(932,179)
\put(155,179){\makebox(0,0)[r]{15}}
\thicklines \path(177,371)(197,371)
\thicklines \path(952,371)(932,371)
\put(155,371){\makebox(0,0)[r]{20}}
\thicklines \path(177,562)(197,562)
\thicklines \path(952,562)(932,562)
\put(155,562){\makebox(0,0)[r]{25}}
\thicklines \path(177,754)(197,754)
\thicklines \path(952,754)(932,754)
\put(155,754){\makebox(0,0)[r]{30}}
\thicklines \path(177,945)(197,945)
\thicklines \path(952,945)(932,945)
\put(155,945){\makebox(0,0)[r]{35}}
\thicklines \path(177,179)(177,199)
\thicklines \path(177,945)(177,925)
\put(177,134){\makebox(0,0){0}}
\thicklines \path(371,179)(371,199)
\thicklines \path(371,945)(371,925)
\put(371,134){\makebox(0,0){5}}
\thicklines \path(565,179)(565,199)
\thicklines \path(565,945)(565,925)
\put(565,134){\makebox(0,0){10}}
\thicklines \path(758,179)(758,199)
\thicklines \path(758,945)(758,925)
\put(758,134){\makebox(0,0){15}}
\thicklines \path(952,179)(952,199)
\thicklines \path(952,945)(952,925)
\put(952,134){\makebox(0,0){20}}
\thicklines \path(177,179)(952,179)(952,945)(177,945)(177,179)
\put(25,658){\makebox(0,0)[l]{\shortstack{$\langle \varphi \rangle$ ($^\circ$)}}}
\put(564,50){\makebox(0,0){$\Omega$ (Hz)}}
\put(875,562){\makebox(0,0){\bf (b)}}
\thinlines
\put(775,371){\makebox(0,0)[r]{experiment}}
\dottedline{12}(797,371)(905,371)
\put(196,719){\circle*{12}}
\put(197,731){\circle*{12}}
\put(198,685){\circle*{12}}
\put(200,696){\circle*{12}}
\put(200,684){\circle*{12}}
\put(201,686){\circle*{12}}
\put(203,399){\circle*{12}}
\put(204,374){\circle*{12}}
\put(207,243){\circle*{12}}
\put(211,317){\circle*{12}}
\put(214,408){\circle*{12}}
\put(221,584){\circle*{12}}
\put(226,569){\circle*{12}}
\put(231,568){\circle*{12}}
\put(235,578){\circle*{12}}
\put(242,536){\circle*{12}}
\put(247,554){\circle*{12}}
\put(258,616){\circle*{12}}
\put(267,633){\circle*{12}}
\put(271,644){\circle*{12}}
\put(305,615){\circle*{12}}
\put(311,615){\circle*{12}}
\put(324,642){\circle*{12}}
\put(333,784){\circle*{12}}
\put(344,672){\circle*{12}}
\put(357,665){\circle*{12}}
\put(375,796){\circle*{12}}
\put(389,754){\circle*{12}}
\put(402,735){\circle*{12}}
\put(479,765){\circle*{12}}
\put(503,690){\circle*{12}}
\put(550,777){\circle*{12}}
\put(907,802){\circle*{12}}
\put(851,371){\circle*{12}}
\dottedline{12}(196,719)(197,731)(198,685)(200,696)(200,684)(201,686)(203,399)
\dottedline{12}(203,399)(204,374)(207,243)(211,317)(214,408)(221,584)(226,569)
\dottedline{12}(226,569)(231,568)(235,578)(242,536)(247,554)(258,616)(267,633)
\dottedline{12}(267,633)(271,644)(305,615)(311,615)(324,642)(333,784)(344,672)
\dottedline{12}(344,672)(357,665)(375,796)(389,754)(402,735)(479,765)(503,690)
\dottedline{12}(503,690)(550,777)(907,802)

\thicklines
\put(775,326){\makebox(0,0)[r]{simulation}}
\thinlines \path(797,326)(905,326)
\thinlines \path(196,568)(196,568)(216,572)(255,583)(293,600)(332,629)(371,663)(390,681)(410,704)(429,729)(448,754)(468,758)(487,758)(565,758)(661,758)(758,758)(855,763)
\put(196,568){\circle{18}}
\put(216,572){\circle{18}}
\put(255,583){\circle{18}}
\put(293,600){\circle{18}}
\put(332,629){\circle{18}}
\put(371,663){\circle{18}}
\put(390,681){\circle{18}}
\put(410,704){\circle{18}}
\put(429,729){\circle{18}}
\put(448,754){\circle{18}}
\put(468,758){\circle{18}}
\put(487,758){\circle{18}}
\put(565,758){\circle{18}}
\put(661,758){\circle{18}}
\put(758,758){\circle{18}}
\put(855,763){\circle{18}}
\put(851,326){\circle{18}}
\end{picture} }
  \caption{}
  \label{fig: one_part2}
\end{figure}

\begin{figure}[htb]
  \hbox{\hspace{-0.7cm}% GNUPLOT: LaTeX picture using EEPIC macros
\setlength{\unitlength}{0.225pt}
\begin{picture}(1049,900)(0,50)
\small
\thicklines \path(177,179)(197,179)
\thicklines \path(952,179)(932,179)
\put(155,179){\makebox(0,0)[r]{0}}
\thicklines \path(177,371)(197,371)
\thicklines \path(952,371)(932,371)
\put(155,371){\makebox(0,0)[r]{20}}
\thicklines \path(177,562)(197,562)
\thicklines \path(952,562)(932,562)
\put(155,562){\makebox(0,0)[r]{40}}
\thicklines \path(177,754)(197,754)
\thicklines \path(952,754)(932,754)
\put(155,754){\makebox(0,0)[r]{60}}
\thicklines \path(177,945)(197,945)
\thicklines \path(952,945)(932,945)
\put(155,945){\makebox(0,0)[r]{80}}
\thicklines \path(177,179)(177,199)
\thicklines \path(177,945)(177,925)
\put(177,134){\makebox(0,0){0.6}}
\thicklines \path(371,179)(371,199)
\thicklines \path(371,945)(371,925)
\put(371,134){\makebox(0,0){0.7}}
\thicklines \path(564,179)(564,199)
\thicklines \path(564,945)(564,925)
\put(564,134){\makebox(0,0){0.8}}
\thicklines \path(758,179)(758,199)
\thicklines \path(758,945)(758,925)
\put(758,134){\makebox(0,0){0.9}}
\thicklines \path(952,179)(952,199)
\thicklines \path(952,945)(952,925)
\put(952,134){\makebox(0,0){1}}
\thicklines \path(177,179)(952,179)(952,945)(177,945)(177,179)
\put(10,658){\makebox(0,0)[l]{\shortstack{$\Omega_c$ (Hz)}}}
\put(564,50){\makebox(0,0){$\mu_w$}}
\put(228,849){\makebox(0,0){I}}
\put(337,849){\makebox(0,0){II}}
\put(673,849){\makebox(0,0){III}}
\put(855,754){\makebox(0,0){\bf (a)}}
\thinlines
\path(952,320)(952,320)(855,321)(758,324)(661,325)(565,325)(468,329)(448,329)(429,330)(410,330)(400,331)(390,394)(380,397)(371,414)(361,426)(351,461)(342,509)(332,423)(322,461)(313,557)(303,586)(293,720)(284,873)
\put(952,320){\circle{18}}
\put(855,321){\circle{18}}
\put(758,324){\circle{18}}
\put(661,325){\circle{18}}
\put(565,325){\circle{18}}
\put(468,329){\circle{18}}
\put(448,329){\circle{18}}
\put(429,330){\circle{18}}
\put(410,330){\circle{18}}
\put(400,331){\circle{18}}
\put(390,394){\circle{18}}
\put(380,397){\circle{18}}
\put(371,414){\circle{18}}
\put(361,426){\circle{18}}
\put(351,461){\circle{18}}
\put(342,509){\circle{18}}
\put(332,423){\circle{18}}
\put(322,461){\circle{18}}
\put(313,557){\circle{18}}
\put(303,586){\circle{18}}
\put(293,720){\circle{18}}
\put(284,873){\circle{18}}
%\put(279,179){\circle{18}}
%\put(279,945){\circle{18}}
%\put(395,179){\circle{18}}
%\put(395,945){\circle{18}}
\dottedline{12}(279,179)(279,945)
\dottedline{12}(395,179)(395,945)
\end{picture} \hfill \hspace{-0.7cm}\input few_f3b }
  \caption{}
  \label{fig: one_part3}
\end{figure}

\begin{figure}[htb]
  \hbox{% GNUPLOT: LaTeX picture using EEPIC macros
\setlength{\unitlength}{0.22pt}
\begin{picture}(960,900)(70,50)
%\tenrm
\thicklines \path(199,179)(219,179)
\thicklines \path(952,179)(932,179)
\put(177,179){\makebox(0,0)[r]{-15}}
\thicklines \path(199,307)(219,307)
\thicklines \path(952,307)(932,307)
\put(177,307){\makebox(0,0)[r]{-10}}
\thicklines \path(199,434)(219,434)
\thicklines \path(952,434)(932,434)
\put(177,434){\makebox(0,0)[r]{-5}}
\thicklines \path(199,562)(219,562)
\thicklines \path(952,562)(932,562)
\put(177,562){\makebox(0,0)[r]{0}}
\thicklines \path(199,690)(219,690)
\thicklines \path(952,690)(932,690)
\put(177,690){\makebox(0,0)[r]{5}}
\thicklines \path(199,817)(219,817)
\thicklines \path(952,817)(932,817)
\put(177,817){\makebox(0,0)[r]{10}}
\thicklines \path(199,945)(219,945)
\thicklines \path(952,945)(932,945)
\put(177,945){\makebox(0,0)[r]{15}}
\thicklines \path(199,179)(199,199)
\thicklines \path(199,945)(199,925)
\put(199,134){\makebox(0,0){0}}
\thicklines \path(387,179)(387,199)
\thicklines \path(387,945)(387,925)
\put(387,134){\makebox(0,0){$\pi/8$}}
\thicklines \path(576,179)(576,199)
\thicklines \path(576,945)(576,925)
\put(576,134){\makebox(0,0){$\pi/4$}}
\thicklines \path(764,179)(764,199)
\thicklines \path(764,945)(764,925)
\put(764,134){\makebox(0,0){$3\pi/8$}}
\thicklines \path(952,179)(952,199)
\thicklines \path(952,945)(952,925)
\put(952,134){\makebox(0,0){$\pi/2$}}
\thicklines \path(199,179)(952,179)(952,945)(199,945)(199,179)
\put(60,562){\makebox(0,0)[l]{\shortstack{$\dot{\varphi}$}}}
\put(575,40){\makebox(0,0){$\varphi$}}
\thicklines
\ifx\color\undefined\thinlines\else\color{red}\fi
%\put(778,903){\makebox(0,0)[r]{$\Omega$=1 Hz}}
%\path(800,903)(908,903)
\path(203,588)(203,588)(209,587)(215,587)(222,587)(228,587)(234,587)(240,587)(246,587)(252,587)(258,587)(265,587)(271,587)(277,587)(283,587)(289,587)(295,587)(301,587)(307,587)(313,587)(319,587)(325,587)(331,587)(337,587)(343,587)(349,587)(355,587)(361,587)(367,586)(373,586)(379,586)(385,586)(391,586)(396,586)(402,586)(408,586)(414,586)(420,586)(426,586)(432,586)(437,586)(443,586)(449,586)(455,586)(461,586)(467,585)(473,584)(478,581)(482,578)(485,575)(488,570)
\path(488,570)(489,566)(489,561)(488,557)(487,552)(484,548)(480,545)(476,542)(471,540)(465,538)(459,538)(453,538)(448,540)(442,542)(438,544)(434,548)(432,552)(429,556)(428,560)(428,564)(429,569)(432,573)(435,577)(438,580)(443,583)(448,584)(454,586)(460,586)(466,585)(471,584)(476,582)
\ifx\color\undefined\thicklines\else\color{green}\fi
%\put(778,858){\makebox(0,0)[r]{2 Hz}}
%\path(800,858)(908,858)
\path(203,598)(203,598)(215,613)(227,613)(240,613)(252,613)(264,613)(277,613)(289,613)(301,612)(314,612)(326,612)(338,612)(350,612)(362,612)(375,612)(386,612)(399,612)(411,612)(423,612)(435,612)(447,612)(459,612)(471,611)(483,609)(494,605)(504,600)(512,593)(519,584)(523,575)(525,566)(524,556)(522,546)(517,537)(510,529)(501,523)(491,518)(480,514)(468,513)(456,513)(444,515)(433,518)(423,523)(414,529)(407,537)(402,545)(399,553)(398,562)(399,571)(402,579)(407,587)
\path(407,587)(414,594)(423,601)(433,606)(444,609)(456,611)(468,611)(479,610)(491,606)(501,601)(510,595)
\ifx\color\undefined\Thicklines\else\color{blue}\fi
%\put(778,813){\makebox(0,0)[r]{5 Hz}}
%\path(800,813)(908,813)
\path(203,598)(203,598)(217,634)(238,669)(268,689)(299,689)(330,689)(360,689)(391,688)(422,688)(453,688)(483,688)(514,687)(544,681)(571,670)(596,655)(616,636)(632,613)(641,588)(643,561)(640,535)(631,510)(615,487)(595,468)(570,453)(542,443)(512,437)(482,436)(452,440)(422,447)(396,458)(372,472)(352,488)(336,506)(325,525)(319,545)(317,566)(320,586)(329,606)(341,625)(359,642)(381,657)(405,670)(432,680)(463,686)(493,688)(523,685)
\thicklines
%\ifx\color\undefined\thinlines\else\color{cyan}\fi
%\put(778,768){\makebox(0,0)[r]{11 Hz}}
%\path(800,768)(908,768)
%\ifx\color\undefined
\dottedline{12}(203,598)(203,598)(217,634)(238,669)(269,703)(307,736)(353,765)(406,791)(464,812)(525,816)(587,815)(648,811)(708,798)(762,775)(811,742)(849,701)(877,652)(893,598)(894,542)(883,488)(859,437)(823,393)(777,358)(724,332)(667,316)(606,309)(544,312)(484,322)(427,339)(376,361)(330,387)(291,416)(259,448)(236,481)(220,515)(213,549)(214,584)(223,618)(241,652)(267,685)(300,716)(341,744)(389,770)(442,791)(499,806)(560,814)(622,814)
%\else
%\path(203,598)(203,598)(217,634)(238,669)(269,703)(307,736)(353,765)(406,791)(464,812)(525,816)(587,815)(648,811)(708,798)(762,775)(811,742)(849,701)(877,652)(893,598)(894,542)(883,488)(859,437)(823,393)(777,358)(724,332)(667,316)(606,309)(544,312)(484,322)(427,339)(376,361)(330,387)(291,416)(259,448)(236,481)(220,515)(213,549)(214,584)(223,618)(241,652)(267,685)(300,716)(341,744)(389,770)(442,791)(499,806)(560,814)(622,814)
%\fi
%\ifx\color\undefined\thinlines\else\color{magenta}\fi
%\put(778,723){\makebox(0,0)[r]{20 Hz}}
%\path(800,723)(908,723)
%\ifx\color\undefined
\drawline[-50](201,584)(205,605)(213,627)(224,648)(238,669)(256,690)(276,710)(299,729)(325,748)(353,765)(384,781)(417,796)(452,808)(488,820)(527,828)(565,833)(606,835)(645,833)(684,830)(723,822)(760,810)(795,794)(827,774)(857,751)(883,725)(904,696)(921,664)(934,630)(941,595)(944,559)(940,523)(932,488)(919,455)(901,423)(878,394)(852,368)(822,346)(790,327)(754,312)(717,302)(678,294)(638,289)(599,289)(559,291)(520,299)(483,307)(446,318)(412,330)(378,345)(348,362)(320,379)(294,398)(272,417)(253,437)(236,458)(222,479)(212,500)(204,522)(200,544)(199,566)
\end{picture} \hfill % GNUPLOT: LaTeX picture using EEPIC macros
\setlength{\unitlength}{0.22pt}
\begin{picture}(960,900)(60,50)
%\tenrm
\thicklines \path(221,270)(241,270)
\thicklines \path(952,270)(932,270)
\put(199,270){\makebox(0,0)[r]{}}
\thicklines \path(221,366)(241,366)
\thicklines \path(952,366)(932,366)
\put(199,366){\makebox(0,0)[r]{-0.5}}
\thicklines \path(221,463)(241,463)
\thicklines \path(952,463)(932,463)
\put(199,463){\makebox(0,0)[r]{}}
\thicklines \path(221,559)(241,559)
\thicklines \path(952,559)(932,559)
\put(199,559){\makebox(0,0)[r]{0}}
\thicklines \path(221,656)(241,656)
\thicklines \path(952,656)(932,656)
\put(199,656){\makebox(0,0)[r]{}}
\thicklines \path(221,752)(241,752)
\thicklines \path(952,752)(932,752)
\put(199,752){\makebox(0,0)[r]{0.5}}
\thicklines \path(221,849)(241,849)
\thicklines \path(952,849)(932,849)
\put(199,849){\makebox(0,0)[r]{}}
\thicklines \path(221,945)(241,945)
\thicklines \path(952,945)(932,945)
\put(199,945){\makebox(0,0)[r]{1}}
\thicklines \path(221,179)(221,199)
\thicklines \path(221,945)(221,925)
\put(221,134){\makebox(0,0){0}}
\thicklines \path(404,179)(404,199)
\thicklines \path(404,945)(404,925)
\put(404,134){\makebox(0,0){$\pi/8$}}
\thicklines \path(587,179)(587,199)
\thicklines \path(587,945)(587,925)
\put(587,134){\makebox(0,0){$\pi/4$}}
\thicklines \path(769,179)(769,199)
\thicklines \path(769,945)(769,925)
\put(769,134){\makebox(0,0){$3\pi/8$}}
\thicklines \path(952,179)(952,199)
\thicklines \path(952,945)(952,925)
\put(952,134){\makebox(0,0){$\pi/2$}}
\thicklines \path(221,179)(952,179)(952,945)(221,945)(221,179)
\put(80,562){\makebox(0,0)[l]{\shortstack{\large $\frac{\dot{\varphi}}{\Omega}$}}}
\put(586,40){\makebox(0,0){$\varphi$}}
\thicklines
\ifx\color\undefined\thinlines\else\color{red}\fi
%\put(778,903){\makebox(0,0)[r]{$\Omega$=1 Hz}}
%\path(800,903)(908,903)
\path(225,945)(225,945)(231,945)(237,944)(243,943)(249,943)(255,942)(261,942)(267,941)(273,940)(279,940)(285,939)(291,938)(296,938)(302,937)(308,937)(314,936)(320,936)(326,935)(331,935)(338,934)(344,933)(349,933)(355,932)(361,932)(366,931)(373,931)(379,930)(384,930)(390,929)(396,928)(401,928)(407,928)(413,927)(418,926)(424,926)(430,925)(435,925)(441,925)(447,924)(452,923)(458,923)(464,923)(470,922)(475,921)(481,910)(487,887)(491,851)(496,804)(499,749)(501,686)
\path(501,686)(503,618)(503,548)(502,479)(500,412)(498,351)(494,298)(490,255)(485,223)(479,204)(474,197)(468,203)(463,222)(457,253)(453,294)(449,343)(447,401)(444,463)(444,529)(444,596)(444,661)(447,723)(450,780)(453,829)(458,869)(463,899)(469,916)(474,921)(480,914)(485,893)(490,860)
\ifx\color\undefined\thicklines\else\color{green}\fi
%\put(778,858){\makebox(0,0)[r]{2 Hz}}
%\path(800,858)(908,858)
\path(225,833)(225,833)(237,945)(249,944)(261,943)(273,943)(285,942)(296,942)(309,941)(320,940)(332,940)(344,939)(356,939)(368,938)(379,938)(392,937)(403,937)(415,936)(426,936)(439,935)(450,935)(462,934)(474,933)(485,930)(496,914)(507,886)(517,844)(525,791)(531,729)(535,660)(538,587)(537,513)(535,440)(530,373)(523,313)(514,263)(504,225)(494,199)(483,186)(470,187)(459,202)(448,228)(439,265)(430,312)(423,367)(418,427)(415,492)(414,558)(415,625)(418,689)(423,750)
\path(423,750)(430,805)(439,852)(448,889)(459,916)(470,931)(482,932)(493,920)(504,895)(514,857)(523,807)
\ifx\color\undefined\Thicklines\else\color{blue}\fi
%\put(778,813){\makebox(0,0)[r]{5 Hz}}
%\path(800,813)(908,813)
\path(225,669)(225,669)(238,777)(259,884)(288,944)(318,943)(348,943)(378,942)(408,942)(438,941)(467,941)(497,940)(526,936)(556,918)(582,886)(606,840)(626,782)(641,713)(650,637)(652,557)(649,477)(640,401)(625,333)(605,275)(581,230)(554,199)(525,182)(496,179)(466,189)(438,212)(412,245)(389,287)(370,335)(354,390)(344,448)(337,509)(336,571)(339,632)(347,692)(359,749)(376,802)(397,848)(421,887)(448,916)(477,934)(506,940)(535,932)
\thicklines
%\ifx\color\undefined\thinlines\else\color{cyan}\fi
%\put(778,768){\makebox(0,0)[r]{11 Hz}}
%\path(800,768)(908,768)
%\ifx\color\undefined
\dottedline{12}(225,609)(225,609)(238,659)(259,707)(289,754)(326,798)(370,839)(422,874)(478,903)(539,924)(603,934)(667,931)(729,916)(789,886)(841,842)(885,784)(917,714)(937,637)(944,555)(936,474)(915,397)(881,329)(836,272)(783,229)(724,201)(660,186)(596,186)(533,197)(472,218)(416,248)(366,284)(322,325)(285,369)(257,416)(236,465)(224,515)(221,564)(226,614)(240,664)(262,712)(292,759)
%\else
%\path(225,609)(225,609)(238,659)(259,707)(289,754)(326,798)(370,839)(422,874)(478,903)(539,924)(603,934)(667,931)(729,916)(789,886)(841,842)(885,784)(917,714)(937,637)(944,555)(936,474)(915,397)(881,329)(836,272)(783,229)(724,201)(660,186)(596,186)(533,197)(472,218)(416,248)(366,284)(322,325)(285,369)(257,416)(236,465)(224,515)(221,564)(226,614)(240,664)(262,712)(292,759)
%\fi
%\ifx\color\undefined\thinlines\else\color{magenta}\fi
%\put(778,723){\makebox(0,0)[r]{20 Hz}}
%\path(800,723)(908,723)
%\ifx\color\undefined
\drawline[-50](223,576)(227,592)(235,608)(246,625)(259,640)(276,656)(295,671)(318,686)(343,700)(370,713)(401,725)(432,736)(466,746)(502,754)(539,760)(577,764)(616,765)(654,765)(692,762)(729,755)(766,747)(799,735)(831,720)(859,703)(885,682)(906,660)(922,636)(934,611)(941,584)(944,557)(941,530)(933,504)(920,478)(902,454)(881,433)(855,413)(826,396)(794,382)(760,371)(724,362)(686,356)(647,354)(609,353)(570,355)(533,360)(496,366)(461,374)(427,384)(395,396)(366,408)(339,421)(314,435)(292,450)(273,465)(257,481)(244,497)(233,513)(226,529)(222,546)(221,562)
\end{picture} }
  \caption{}
  \label{fig: one_part4}
\end{figure}

\begin{figure}[htb]
  \hbox{%LaTeX picture with EEPIC extensions
\setlength{\unitlength}{0.155pt}
\begin{picture}(850,900)(0,0)
\small
\put(100,700){\makebox(0,0){(a)}}
\thicklines \put(449,418){\circle{738}}
\thinlines
\put(722,230){\shade\circle{ 73.8}}
\put(474, 87){\shade\circle{ 73.8}}
\put(547,101){\shade\circle{ 73.8}}
\put(674,175){\shade\circle{ 73.8}}
\put(614,131){\shade\circle{ 73.8}}
\end{picture} %LaTeX picture with EEPIC extensions
\setlength{\unitlength}{0.155pt}
\begin{picture}(850,900)(0,0)
\small
\put(100,700){\makebox(0,0){(b)}}
\thicklines \put(449,418){\circle{738}}
\thinlines
\put(364,190){\shade\circle{ 73.8}}
\put(464,161){\shade\circle{ 73.8}}
\put(682,328){\shade\circle{ 73.8}}
\put(650,261){\shade\circle{ 73.8}}
\put(395, 91){\shade\circle{ 73.8}}
\put(536,173){\shade\circle{ 73.8}}
\put(772,363){\shade\circle{ 73.8}}
\put(323,112){\shade\circle{ 73.8}}
\put(609,128){\shade\circle{ 73.8}}
\put(670,171){\shade\circle{ 73.8}}
\put(467, 87){\shade\circle{ 73.8}}
\put(540, 99){\shade\circle{ 73.8}}
\put(755,291){\shade\circle{ 73.8}}
\put(604,204){\shade\circle{ 73.8}}
\put(716,228){\shade\circle{ 73.8}}
\end{picture} %LaTeX picture with EEPIC extensions
\setlength{\unitlength}{0.155pt}
\begin{picture}(850,900)(0,0)
\small
\put(100,700){\makebox(0,0){(c)}}
\thicklines \put(449,418){\circle{738}}
\thinlines
\put(769,331){\shade\circle{ 73.8}}
\put(274,699){\shade\circle{ 73.8}}
\put(737,582){\shade\circle{ 73.8}}
\put(729,240){\shade\circle{ 73.8}}
\put(257,148){\shade\circle{ 73.8}}
\put(170,239){\shade\circle{ 73.8}}
\put(781,418){\shade\circle{ 73.8}}
\put(379,742){\shade\circle{ 73.8}}
\put(766,514){\shade\circle{ 73.8}}
\put(211,648){\shade\circle{ 73.8}}
\put(118,400){\shade\circle{ 73.8}}
\put(508, 92){\shade\circle{ 73.8}}
\put(609,128){\shade\circle{ 73.8}}
\put(575,725){\shade\circle{ 73.8}}
\put(410, 89){\shade\circle{ 73.8}}
\put(131,326){\shade\circle{ 73.8}}
\put(483,748){\shade\circle{ 73.8}}
\put(692,644){\shade\circle{ 73.8}}
\put(150,562){\shade\circle{ 73.8}}
\put(125,489){\shade\circle{ 73.8}}
\end{picture} }
  \caption{}
  \label{fig: few_part1}
\end{figure}

\begin{figure}[htb]
  \hbox{\hspace{-0.7cm}% GNUPLOT: LaTeX picture using EEPIC macros
\setlength{\unitlength}{0.225pt}
\begin{picture}(1049,900)(0,50)
\small
\thicklines \path(177,179)(197,179)
\thicklines \path(952,179)(932,179)
\put(155,179){\makebox(0,0)[r]{0}}
\thicklines \path(177,307)(197,307)
\thicklines \path(952,307)(932,307)
\put(155,307){\makebox(0,0)[r]{}}
\thicklines \path(177,434)(197,434)
\thicklines \path(952,434)(932,434)
\put(155,434){\makebox(0,0)[r]{20}}
\thicklines \path(177,562)(197,562)
\thicklines \path(952,562)(932,562)
\put(155,562){\makebox(0,0)[r]{}}
\thicklines \path(177,690)(197,690)
\thicklines \path(952,690)(932,690)
\put(155,690){\makebox(0,0)[r]{40}}
\thicklines \path(177,817)(197,817)
\thicklines \path(952,817)(932,817)
\put(155,817){\makebox(0,0)[r]{}}
\thicklines \path(177,945)(197,945)
\thicklines \path(952,945)(932,945)
\put(155,945){\makebox(0,0)[r]{60}}
\thicklines \path(177,179)(177,199)
\thicklines \path(177,945)(177,925)
\put(177,134){\makebox(0,0){0}}
\thicklines \path(306,179)(306,199)
\thicklines \path(306,945)(306,925)
\put(306,134){\makebox(0,0){}}
\thicklines \path(435,179)(435,199)
\thicklines \path(435,945)(435,925)
\put(435,134){\makebox(0,0){1}}
\thicklines \path(565,179)(565,199)
\thicklines \path(565,945)(565,925)
\put(565,134){\makebox(0,0){}}
\thicklines \path(694,179)(694,199)
\thicklines \path(694,945)(694,925)
\put(694,134){\makebox(0,0){2}}
\thicklines \path(823,179)(823,199)
\thicklines \path(823,945)(823,925)
\put(823,134){\makebox(0,0){}}
\thicklines \path(952,179)(952,199)
\thicklines \path(952,945)(952,925)
\put(952,134){\makebox(0,0){3}}
\thicklines \path(177,179)(952,179)(952,945)(177,945)(177,179)
\put(25,562){\makebox(0,0)[l]{\shortstack{$\langle\varphi\rangle$ ($^\circ$)}}}
\put(564,67){\makebox(0,0){$t$ (s)}}
\put(823,754){\makebox(0,0){\bf (a)}}
\put(801,881){\makebox(0,0)[r]{simulation}}
\thinlines \path(823,881)(931,881)
\thinlines \path(177,179)(177,179)(185,208)(192,292)(200,420)(207,569)(215,713)(223,819)(230,863)(238,832)(245,732)(253,594)(260,434)(268,318)(276,242)(283,216)(291,246)(298,325)(306,441)(314,574)(321,698)(329,789)(336,826)(344,801)(351,720)(359,600)(367,467)(374,345)(382,255)(389,210)(397,214)(405,269)(412,366)(420,492)(427,624)(435,738)(442,810)(450,822)(458,774)(465,674)(473,546)(480,415)(488,304)(496,231)(503,204)(511,210)(518,243)(526,301)(533,376)(541,460)(549,542)
\thinlines \path(549,542)(556,609)(564,651)(571,660)(579,633)(587,577)(594,501)(602,417)(609,336)(617,269)(624,223)(632,204)(640,212)(647,248)(655,308)(662,384)(670,468)(678,549)(685,614)(693,653)(700,658)(708,629)(715,571)(723,493)(731,409)(738,330)(746,264)(753,221)(761,204)(769,215)(776,253)(784,314)(791,392)(799,476)(807,555)(814,618)(822,654)(829,656)(837,624)(844,564)(852,486)(860,402)(867,323)(875,260)(882,219)(890,204)(898,218)(905,258)(913,321)(920,399)(928,483)
\thinlines \path(928,483)(935,562)
\put(801,836){\makebox(0,0)[r]{arctan($\mu_w$)}}
\dottedline{12}(823,836)(931,836)
\dottedline{12}(177,425)(952,425)
\end{picture} \hfill \hspace{-0.7cm}% GNUPLOT: LaTeX picture using EEPIC macros
%\setlength{\unitlength}{0.240900pt}
%\begin{picture}(1500,900)(0,0)
\setlength{\unitlength}{0.225pt}
\begin{picture}(1049,900)(0,50)
%\tenrm
\thicklines \path(177,179)(197,179)
\thicklines \path(952,179)(932,179)
\put(155,179){\makebox(0,0)[r]{0}}
\thicklines \path(177,307)(197,307)
\thicklines \path(952,307)(932,307)
\put(155,307){\makebox(0,0)[r]{}}
\thicklines \path(177,434)(197,434)
\thicklines \path(952,434)(932,434)
\put(155,434){\makebox(0,0)[r]{10}}
\thicklines \path(177,562)(197,562)
\thicklines \path(952,562)(932,562)
\put(155,562){\makebox(0,0)[r]{}}
\thicklines \path(177,690)(197,690)
\thicklines \path(952,690)(932,690)
\put(155,690){\makebox(0,0)[r]{20}}
\thicklines \path(177,817)(197,817)
\thicklines \path(952,817)(932,817)
\put(155,817){\makebox(0,0)[r]{}}
\thicklines \path(177,945)(197,945)
\thicklines \path(952,945)(932,945)
\put(155,945){\makebox(0,0)[r]{30}}
\thicklines \path(177,179)(177,199)
\thicklines \path(177,945)(177,925)
\put(177,134){\makebox(0,0){0.2}}
\thicklines \path(332,179)(332,199)
\thicklines \path(332,945)(332,925)
\put(332,134){\makebox(0,0){0.3}}
\thicklines \path(487,179)(487,199)
\thicklines \path(487,945)(487,925)
\put(487,134){\makebox(0,0){0.4}}
\thicklines \path(642,179)(642,199)
\thicklines \path(642,945)(642,925)
\put(642,134){\makebox(0,0){0.5}}
\thicklines \path(797,179)(797,199)
\thicklines \path(797,945)(797,925)
\put(797,134){\makebox(0,0){0.6}}
\thicklines \path(952,179)(952,199)
\thicklines \path(952,945)(952,925)
\put(952,134){\makebox(0,0){0.7}}
\thicklines \path(177,179)(952,179)(952,945)(177,945)(177,179)
\put(80,562){\makebox(0,0)[l]{\shortstack{$N_c$}}}
\put(564,67){\makebox(0,0){$\mu_w$}}
\put(332,307){\makebox(0,0){\bf (b)}}
\put(775,881){\makebox(0,0)[r]{$\mu=0.00$}}
\thinlines \path(797,881)(905,881)
\thinlines \path(177,542)(177,542)(185,541)(193,540)(200,539)(208,538)(216,536)(224,535)(232,534)(240,533)(247,531)(255,530)(263,529)(271,528)(279,527)(287,525)(294,524)(302,523)(310,522)(318,521)(326,519)(334,518)(341,517)(349,516)(357,515)(365,513)(373,512)(381,511)(388,510)(396,509)(404,508)(412,507)(420,505)(428,504)(435,503)(443,502)(451,501)(459,500)(467,499)(474,497)(482,496)(490,495)(498,494)(506,493)(514,492)(521,491)(529,490)(537,489)(545,488)(553,486)(561,485)
\thinlines \path(561,485)(568,484)(576,483)(584,482)(592,481)(600,480)(608,479)(615,478)(623,477)(631,476)(639,475)(647,474)(655,473)(662,472)(670,471)(678,470)(686,469)(694,468)(701,467)(709,466)(717,465)(725,464)(733,463)(741,462)(748,461)(756,460)(764,459)(772,458)(780,457)(788,456)(795,455)(803,454)(811,453)(819,452)(827,451)(835,450)(842,449)(850,449)(858,448)(866,447)(874,446)(882,445)(889,444)(897,443)(905,442)(913,441)(921,440)(929,440)(936,439)(944,438)(952,437)
\put(775,836){\makebox(0,0)[r]{$0.25$}}
\thinlines \path(797,836)(905,836)
\thinlines \path(797,846)(797,826)
\thinlines \path(905,846)(905,826)
\thinlines \path(293,868)(293,894)
\thinlines \path(283,868)(303,868)
\thinlines \path(283,894)(303,894)
\thinlines \path(313,817)(313,843)
\thinlines \path(303,817)(323,817)
\thinlines \path(303,843)(323,843)
\thinlines \path(332,792)(332,817)
\thinlines \path(322,792)(342,792)
\thinlines \path(322,817)(342,817)
\thinlines \path(351,766)(351,792)
\thinlines \path(341,766)(361,766)
\thinlines \path(341,792)(361,792)
\thinlines \path(371,690)(371,715)
\thinlines \path(361,690)(381,690)
\thinlines \path(361,715)(381,715)
\thinlines \path(390,639)(390,664)
\thinlines \path(380,639)(400,639)
\thinlines \path(380,664)(400,664)
\thinlines \path(410,639)(410,664)
\thinlines \path(400,639)(420,639)
\thinlines \path(400,664)(420,664)
\thinlines \path(429,613)(429,639)
\thinlines \path(419,613)(439,613)
\thinlines \path(419,639)(439,639)
\thinlines \path(448,588)(448,613)
\thinlines \path(438,588)(458,588)
\thinlines \path(438,613)(458,613)
\thinlines \path(468,613)(468,639)
\thinlines \path(458,613)(478,613)
\thinlines \path(458,639)(478,639)
\thinlines \path(487,562)(487,588)
\thinlines \path(477,562)(497,562)
\thinlines \path(477,588)(497,588)
\thinlines \path(526,511)(526,536)
\thinlines \path(516,511)(536,511)
\thinlines \path(516,536)(536,536)
\thinlines \path(565,511)(565,536)
\thinlines \path(555,511)(575,511)
\thinlines \path(555,536)(575,536)
\thinlines \path(603,460)(603,485)
\thinlines \path(593,460)(613,460)
\thinlines \path(593,485)(613,485)
\thinlines \path(642,460)(642,485)
\thinlines \path(632,460)(652,460)
\thinlines \path(632,485)(652,485)
\thinlines \path(681,460)(681,485)
\thinlines \path(671,460)(691,460)
\thinlines \path(671,485)(691,485)
\thinlines \path(720,460)(720,485)
\thinlines \path(710,460)(730,460)
\thinlines \path(710,485)(730,485)
\thinlines \path(758,511)(758,536)
\thinlines \path(748,511)(768,511)
\thinlines \path(748,536)(768,536)
\thinlines \path(797,460)(797,485)
\thinlines \path(787,460)(807,460)
\thinlines \path(787,485)(807,485)
\thinlines \path(836,434)(836,460)
\thinlines \path(826,434)(846,434)
\thinlines \path(826,460)(846,460)
\put(293,881){\circle{18}}
\put(313,830){\circle{18}}
\put(332,805){\circle{18}}
\put(351,779){\circle{18}}
\put(371,702){\circle{18}}
\put(390,651){\circle{18}}
\put(410,651){\circle{18}}
\put(429,626){\circle{18}}
\put(448,600){\circle{18}}
\put(468,626){\circle{18}}
\put(487,575){\circle{18}}
\put(526,524){\circle{18}}
\put(565,524){\circle{18}}
\put(603,473){\circle{18}}
\put(642,473){\circle{18}}
\put(681,473){\circle{18}}
\put(720,473){\circle{18}}
\put(758,524){\circle{18}}
\put(797,473){\circle{18}}
\put(836,447){\circle{18}}
\put(851,836){\circle{18}}
\put(775,791){\makebox(0,0)[r]{$0.50$}}
\thinlines \path(797,791)(905,791)
\thinlines \path(797,801)(797,781)
\thinlines \path(905,801)(905,781)
\thinlines \path(332,817)(332,843)
\thinlines \path(322,817)(342,817)
\thinlines \path(322,843)(342,843)
\thinlines \path(371,741)(371,766)
\thinlines \path(361,741)(381,741)
\thinlines \path(361,766)(381,766)
\thinlines \path(410,639)(410,664)
\thinlines \path(400,639)(420,639)
\thinlines \path(400,664)(420,664)
\thinlines \path(448,588)(448,613)
\thinlines \path(438,588)(458,588)
\thinlines \path(438,613)(458,613)
\thinlines \path(487,536)(487,562)
\thinlines \path(477,536)(497,536)
\thinlines \path(477,562)(497,562)
\thinlines \path(526,460)(526,485)
\thinlines \path(516,460)(536,460)
\thinlines \path(516,485)(536,485)
\thinlines \path(565,460)(565,485)
\thinlines \path(555,460)(575,460)
\thinlines \path(555,485)(575,485)
\thinlines \path(603,409)(603,434)
\thinlines \path(593,409)(613,409)
\thinlines \path(593,434)(613,434)
\thinlines \path(642,358)(642,383)
\thinlines \path(632,358)(652,358)
\thinlines \path(632,383)(652,383)
\thinlines \path(681,358)(681,383)
\thinlines \path(671,358)(691,358)
\thinlines \path(671,383)(691,383)
\thinlines \path(720,358)(720,383)
\thinlines \path(710,358)(730,358)
\thinlines \path(710,383)(730,383)
\thinlines \path(758,307)(758,332)
\thinlines \path(748,307)(768,307)
\thinlines \path(748,332)(768,332)
\thinlines \path(797,307)(797,332)
\thinlines \path(787,307)(807,307)
\thinlines \path(787,332)(807,332)
\thinlines \path(875,307)(875,332)
\thinlines \path(865,307)(885,307)
\thinlines \path(865,332)(885,332)
\put(332,830){\circle*{12}}
\put(371,754){\circle*{12}}
\put(410,651){\circle*{12}}
\put(448,600){\circle*{12}}
\put(487,549){\circle*{12}}
\put(526,473){\circle*{12}}
\put(565,473){\circle*{12}}
\put(603,422){\circle*{12}}
\put(642,371){\circle*{12}}
\put(681,371){\circle*{12}}
\put(720,371){\circle*{12}}
\put(758,319){\circle*{12}}
\put(797,319){\circle*{12}}
\put(875,319){\circle*{12}}
\put(851,791){\circle*{12}}
\put(775,746){\makebox(0,0)[r]{fit}}
\thinlines \dottedline{12}(797,746)(905,746)
\thinlines
\dottedline{12}(303,945)(310,919)(318,892)(326,865)(334,840)(341,817)(349,794)(357,773)(365,753)(373,734)(381,715)(388,698)(396,681)(404,665)(412,650)(420,636)(428,622)(435,609)(443,596)(451,584)(459,572)(467,561)(474,551)(482,540)(490,530)(498,521)(506,512)(514,503)(521,495)(529,487)(537,479)(545,471)(553,464)(561,457)(568,450)(576,444)(584,438)(592,432)(600,426)(608,420)(615,415)(623,409)(631,404)(639,399)(647,394)(655,390)(662,385)(670,381)(678,377)(686,373)(694,369)
\dottedline{12}(694,369)(701,365)(709,361)(717,357)(725,354)(733,350)(741,347)(748,344)(756,341)(764,338)(772,335)(780,332)(788,329)(795,326)(803,323)(811,321)(819,318)(827,316)(835,313)(842,311)(850,309)(858,307)(866,304)(874,302)(882,300)(889,298)(897,296)(905,294)(913,292)(921,291)(929,289)(936,287)(944,285)(952,284)
\end{picture} }
  \caption{}
  \label{fig: few_part2}
\end{figure}

\fi

\cleardoublepage

\end{document}